\pdfoutput=1
\documentclass[a4paper,11pt]{article}

\usepackage{fullpage} 
\usepackage{parskip} 
\usepackage{tikz} 
\usepackage{hyperref}
\usepackage{multirow}
\usepackage{stfloats}
\usepackage[bottom]{footmisc}
\usepackage[margin=2.5cm, headheight=16pt, headsep=0.4in, heightrounded]{geometry}
\usepackage{microtype}
\usepackage{kantlipsum}
\usepackage{bbding}
\usepackage[labelfont=bf]{caption}
\usepackage[font={small}]{caption}
\usepackage{enumerate, xfrac, color, graphicx}
\usepackage{amsmath, amsthm, amssymb, amsfonts, nccmath}
\usepackage{booktabs}
\usepackage{caption}
\usepackage{algorithm}
\usepackage{enumitem}
\usepackage{algpseudocode}
\usepackage{pifont}
\usepackage{fancyhdr}
\usepackage{siunitx}
\usepackage{multicol}
\usepackage{stfloats}
\usepackage{rotating}
\usepackage[bottom]{footmisc}

\usepackage{titlesec}
\titlespacing*{\subsection}
  {0pt}{1\baselineskip}{.5\baselineskip}
  
\newcommand\blfootnote[1]{%
  \begingroup
  \renewcommand\thefootnote{}\footnote{#1}%
  \addtocounter{footnote}{-1}%
  \endgroup
}

\makeatletter
\def\blfootnote{\gdef\@thefnmark{}\@footnotetext}
\makeatother

\usepackage{setspace}
\usepackage{lineno}
\usepackage{csquotes}

\usepackage[utf8]{inputenc}
\usepackage[sorting=none]{biblatex}

\usepackage{notoccite}

\DeclareUnicodeCharacter{FB02}{fl}
\DeclareUnicodeCharacter{0301}{\'{e}}

\newrobustcmd*{\parentexttrack}[1]{%
  \begingroup
  \blx@blxinit
  \blx@setsfcodes
  \blx@bibopenparen#1\blx@bibcloseparen
  \endgroup} 

\usepackage{epstopdf}

\usepackage{todonotes}
\usepackage[authormarkup=none]{changes}

\hypersetup{colorlinks=true, linkcolor={blue}, citecolor={blue}, urlcolor={magenta}}

\title{\vspace{36pt}\textbf{\Large{Planning low-carbon distributed power systems: Evaluating the role of energy storage\blfootnote{Published In: \textit{Energy} 238 (2022): 121668.}\blfootnote{DOI: \url{https://doi.org/10.1016/j.energy.2021.121668}.}\\[30pt]}}}

\author{\normalsize \textbf{Jiachen Mao}\footnote{Corresponding authors. E-mail: \href{mailto:j.mao@alum.mit.edu}{j.mao@alum.mit.edu} (J. Mao); \href{mailto:audunb@mit.edu}{audunb@mit.edu} (A. Botterud).} \hspace{1pt}\footnote{Equal contribution.} \hspace{24pt} \textbf{Mehdi Jafari}\footnotemark[2] \hspace{24pt} \textbf{Audun Botterud}\footnotemark[1]\\[24pt]
\normalsize Laboratory for Information and Decision Systems (LIDS),\\ \normalsize Massachusetts Institute of Technology (MIT), Cambridge, MA 02139, USA\\[12pt]}

\date{\vspace{24pt}}

\begin{document}
\maketitle
\pagestyle{fancy}
\fancyhf{}
\chead{\textsc{\small J. Mao, \hspace{2pt} M. Jafari, \hspace{2pt} \& \hspace{2pt} A. Botterud}}
\cfoot{\vspace{0.5pt} \thepage}
\renewcommand{\headrulewidth}{0pt}
\thispagestyle{fancy}
\thispagestyle{plain}
\definechangesauthor[color=black]{mark}

\renewcommand{\abstractname}{\normalsize Abstract}

\begin{abstract}
\normalsize
\noindent This paper introduces a mathematical formulation of energy storage systems into a generation capacity expansion framework to evaluate the role of energy storage in the decarbonization of distributed power systems. The modeling framework accounts for dynamic charging/discharging efficiencies and maximum cycling powers as well as cycle and calendar degradation of a Li-ion battery system. Results from a small-scale distributed power system indicate that incorporating the dynamic efficiencies and cycling powers of batteries in the generation planning problem does not significantly change the optimal generation portfolio, while adding substantial computational burden. In contrast, accounting for battery degradation leads to substantially different generation expansion outcomes, especially in deep decarbonization scenarios with larger energy storage capacities. Under the assumptions used in this study, it is found that battery energy storage is economically viable for 2020 only under strict carbon emission constraints. In contrast, given the projected technology advances and corresponding cost reductions, battery energy storage exhibits an attractive option to enable deep decarbonization in 2050.\\\\

\noindent \textit{Keywords}: Distributed energy resources; Decarbonization; Battery energy storage; Battery degradation; Renewable energy 
\end{abstract}

\newpage
\section*{Nomenclature}
\footnotesize
\subsubsection*{Input parameters}
\textit{a. Index}\\[6pt]
\noindent
\begin{tabular}{p{2cm}p{13cm}}
$d$ & daily time step, $d = 1, 2, ..., D$ \\
$i$ & piece of the piece-wise linear function for discharge loss, $i = 1, 2, ..., I$ \\
$j$ & piece of the piece-wise linear function for charge loss, $j = 1, 2, ..., J$ \\
$k$ & SOC level (p.u.), $k = 1, 2, ..., K$ \\
$t$ & hourly time step, $t = 1, 2, ..., T$ \\
\multicolumn{2}{c}{}
\end{tabular}\\
\textit{b. Thermal plant}\\[6pt]
\noindent
\begin{tabular}{p{2cm}p{13cm}}
$\textrm{A}_{\textrm{T}}$ & annuity factor of thermal plant, $\textrm{A}_\textrm{T} = \textrm{r}/ [ 1 - (1 + \textrm{r})^{-\textrm{l}_\textrm{T}} ]$ \\
$\textrm{AF}_{\textrm{T}}(t)$ & availability factor of thermal plant at period $t$, p.u. \\
$\textrm{CO}_{2\textrm{ T}}$ & unit CO$_2$ emissions of thermal plant, kg/kWh \\
$\textrm{IF}_{\textrm{T}}$ & fixed investment cost of thermal plant, \$/kW \\
$\textrm{l}_{\textrm{T}}$ & capital recovery period of thermal plant, year\\
$\textrm{FOM}_{\textrm{T}}$ & fixed operation and maintenance (O\&M) cost of thermal plant, \$/(kW year) \\
$\textrm{VOM}_{\textrm{T}}$ & variable O\&M cost of thermal plant, \$/kWh \\
$\textrm{sc}_{\textrm{T}}$ & unit capacity (discrete) of thermal plant, kW \\
\multicolumn{2}{c}{}
\end{tabular}\\
\textit{c. Solar PV}\\[6pt]
\noindent
\begin{tabular}{p{2cm}p{13cm}}
$\textrm{A}_{\textrm{PV}}$ & annuity factor of PV, $\textrm{A}_\textrm{PV} = \textrm{r}/ [ 1 - (1 + \textrm{r})^{-\textrm{l}_\textrm{PV}} ]$ \\
$\textrm{AF}_{\textrm{PV}}(t)$ & availability factor of PV at period $t$, p.u. \\
$\textrm{IF}_{\textrm{PV}}$ & fixed investment cost of PV, \$/kW \\
$\textrm{l}_{\textrm{PV}}$ & capital recovery period of PV, year\\
$\textrm{FOM}_{\textrm{PV}}$ & fixed O\&M cost of PV, \$/(kW year) \\
$\textrm{VOM}_{\textrm{PV}}$ & variable O\&M cost of PV, \$/kWh \\
\multicolumn{2}{c}{}
\end{tabular}\\
\textit{d. Wind}\\[6pt]
\noindent
\begin{tabular}{p{2cm}p{13cm}}
$\textrm{A}_{\textrm{W}}$ & annuity factor of wind energy, $\textrm{A}_\textrm{W} = \textrm{r}/ [ 1 - (1 + \textrm{r})^{-\textrm{l}_\textrm{W}} ]$ \\
$\textrm{AF}_{\textrm{W}}(t)$ & availability factor of wind energy at period $t$, p.u. \\
$\textrm{IF}_{\textrm{W}}$ & fixed investment cost of wind energy, \$/kW \\
$\textrm{l}_{\textrm{W}}$ & capital recovery period of wind energy, year\\
$\textrm{FOM}_{\textrm{W}}$ & fixed O\&M cost of wind energy, \$/(kW year) \\
$\textrm{VOM}_{\textrm{W}}$ & variable O\&M cost of wind energy, \$/kWh \\
\multicolumn{2}{c}{}
\end{tabular}\\
\noindent
\textit{e. Battery}\\[6pt]
\noindent
\begin{tabular}{p{2cm}p{13cm}}
$\textrm{A}_{\textrm{B}}$ & annuity factor of battery, $\textrm{A}_\textrm{B} = \textrm{r}/ [ 1 - (1 + \textrm{r})^{-\textrm{l}_\textrm{B}} ]$ \\
$\textrm{B}(k)$ & partition on SOC percentage \\
$\textrm{EOL}$ & end-of-life criterion, \% \\
$\textrm{IP}_{\textrm{B}}$ & power component investment cost of battery, \$/kW \\
$\textrm{IE}_{\textrm{B}}$ & energy component investment cost of battery, \$/kWh \\
$\textrm{l}_{\textrm{B}}$ & capital recovery period of battery, year\\
$\textrm{N}_\textrm{cycle}$ & cycle life \\
$\textrm{FOM}_{\textrm{B}}$ & fixed O\&M cost of battery, \$/(kW year) \\
$\textrm{VOM}_{\textrm{B}}$ & variable O\&M cost of battery, \$/kWh \\
$\textrm{P}_{\textrm{c}}^{\textrm{max}}$ & maximum charging power (fraction of installed power capacity), p.u. \\
PC & maximum of each charging power piece \\
$\textrm{P}_{\textrm{d}}^{\textrm{max}}$ & maximum discharging power (fraction of installed power capacity), p.u. \\
PD & maximum of each discharging power piece \\
$\alpha_\textrm{c}(k, j)$ & slope of piece $j$ at level $k$ for charging loss in piece-wise linear function \\
$\alpha_\textrm{d}(k, i)$ & slope of piece $i$ at level $k$ for discharging loss in piece-wise linear function \\
$\textrm{SOC}_{\textrm{max}}$ & maximum state of charge, \% \\
$\textrm{SOC}_{\textrm{min}}$ & minimum state of charge, \% \\
$\eta_\textrm{c}$ & fixed charge efficiency, \%\\
$\eta_\textrm{d}$ & fixed discharge efficiency, \%\\
$\lambda$ & wrap-up tolerance, \% \\
$p$ & fraction of cycle degradation in total degradation, \% \\
\multicolumn{2}{c}{}
\end{tabular}\\[24pt]
\textit{f. Other}\\[4pt]
\noindent
\begin{tabular}{p{2cm}p{13cm}}
$\textrm{CO}_{2\textrm{ M}}$ & unit CO$_2$ emissions from the grid, kg/kWh \\
$\textrm{L}(t)$ & hourly local demand at period $t$, kW \\
r & discount rate, \% \\
$\textrm{P}_\textrm{LC}$ & value of lost load (VOLL), \$/kWh \\
$\textrm{P}_\textrm{M}(t)$ & hourly price of energy from the grid at period $t$, \$/kWh \\
$\textrm{TF}$ & time factor, $\textrm{TF} = D / 365$ \\
$\gamma$ & customized carbon emission percentage reduction, \% \\
\multicolumn{2}{c}{}
\end{tabular}\vspace{-9pt}
\subsubsection*{Decision variables / functions}
\textit{a. Thermal plant}\\[4pt]
\noindent
\begin{tabular}{p{2cm}p{13cm}}
$C_{\textrm{T}}$ & total cost of thermal plant, \$/year \\
$x$ & number of installed thermal plant units \\
$g_{\textrm{T}}(t)$ & output power of thermal plant at period $t$, kW \\
$I_{\textrm{T}}$ & annualized investment cost of thermal plant, \$/year \\
$OM_{\textrm{T}}$ & O\&M cost of thermal plant, \$/year \\
\multicolumn{2}{c}{}
\end{tabular}\\
\textit{b. Solar PV}\\[4pt]
\noindent
\begin{tabular}{p{2cm}p{13cm}}
$C_{\textrm{PV}}$ & total cost of PV, \$/year \\
$c_{\textrm{PV}}$ & installed capacity of PV, kW \\
$g_{\textrm{PV}}(t)$ & output power of PV at period $t$, kW \\
$I_{\textrm{PV}}$ & annualized investment cost of PV, \$/year \\
$OM_{\textrm{PV}}$ & O\&M cost of PV, \$/year \\
\multicolumn{2}{c}{}
\end{tabular}\\
\noindent
\textit{c. Wind}\\[4pt]
\noindent
\begin{tabular}{p{2cm}p{13cm}}
$C_{\textrm{W}}$ & total cost of wind energy, \$/year \\
$c_{\textrm{W}}$ & installed capacity of wind energy, kW \\
$g_{\textrm{W}}(t)$ & output power of wind energy at period $t$, kW \\
$I_{\textrm{W}}$ & annualized investment cost of wind energy, \$/year \\
$OM_{\textrm{W}}$ & O\&M cost of wind energy, \$/year \\
\multicolumn{2}{c}{}
\end{tabular}\\
\textit{d. Battery}\\[4pt]
\noindent
\begin{tabular}{p{2cm}p{13cm}}
$C_{\textrm{B}}$ & cost of battery, \$/year \\
$c_{\textrm{B}}$ & installed power capacity of battery, kW \\
$E_{\textrm{B}}$ & installed energy capacity of battery, kWh \\
$E_{\textrm{BD}}(d)$ & battery capacity on day $d$ considering degradation, kWh \\
$E_{\textrm{BL}}$ & battery capacity loss, kWh \\
$e_{\textrm{Bc}}(t)$ & battery power input at period $t$, kW \\
$e_{\textrm{Bd}}(t)$ & battery power output at period $t$, kW \\
$I_{\textrm{B}}$ & annualized investment cost of battery, \$/year \\
$OM_{\textrm{B}}$ & O\&M cost of battery, \$/year \\
$P_{\textrm{c}}(t)$ & charging power at period $t$, kW \\
$P_{\textrm{c}}^\textrm{loss} (t)$ & charging losses at period $t$, kW \\
$P_{\textrm{d}}(t)$ & discharging power at period $t$, kW \\
$P_{\textrm{d}}^\textrm{loss} (t)$ & discharging losses at period $t$, kW \\
$SOE(0)$ & initial state of energy, kWh \\
$SOE(t)$ & state of energy at period $t$, kWh \\
$u(t, k)$ & activation function for SOC level $k$ at period $t$ \\
$v_\textrm{c} (k, j, t)$& charge amount of piece $j$ for SOC level $k$ at period $t$, kW \\
$v_\textrm{d} (k, i, t)$& discharge amount of piece $i$ for SOC level $k$ at period $t$, kW \\
$w(t)$ & binary variable, 1 if battery is charging at period $t$ \\
$\beta (t)$ & fraction of battery energy used at period $t$, \% \\
\multicolumn{2}{c}{}
\end{tabular}\\
\textit{e. Other}\\[4pt]
\noindent
\begin{tabular}{p{2cm}p{13cm}}
$C_{\textrm{total}}$ & total cost of the grid, \$/year \\
$C_{\textrm{LC}}$ & cost of load curtailment, \$/year \\
$C_{\textrm{M}}$ & cost of energy purchase from the grid, \$/year \\
$c(t)$ & curtailed load at period $t$, kW \\
$g_{\textrm{M}}(t)$ & power supply from the grid at period $t$, kW\\
\multicolumn{2}{c}{}
\end{tabular}

\newpage

\normalsize
\section{Introduction}

Vast resources are being consumed to drive our modern life, which has generated profound energy and environmental problems at local and global scales. In response to mitigating these challenges, most nations are committed to a dramatic reduction in their greenhouse gas (GHG) emissions \cite{UNFCCC}. Clearly, any serious effort to pursue this agenda will necessarily require a transfiguration of the electric power sector. Desirable reductions in the GHG emissions can be achieved largely by substantial cuts in the use of fossil fuels. Power producers are moving to exploit renewable resources for supplying energy in both large centralized generation and small distributed systems \cite{kassakian2011future}. Aside from the social and economic concerns that such a shift creates, the increasing penetration of renewable and distributed energy resources (DERs) will inevitably introduce many technological challenges. Zia et al. \cite{zia2018} reviewed methods and solutions for the DER management in the power systems and showed that DERs will play a substantial role in the future. DERs can enhance the resilience of the distribution systems through microgrid formation \cite{GILANI2020} and enable demand side response, including through generation and storage, by end users \cite{ROLDAN2019}. However, these changes can alter the operation constraints in distribution systems \cite{SANI2018} and new operation regulation schemes will be required under high penetration of DERs \cite{eid2016}.

An energy storage system (ESS), such as lithium-ion batteries, could play an important role toward this deep decarbonization paradigm \cite{jafari2020}. Essentially, one can rely on the use of ESS to match the generation at one time instant and/or location with the demand at another time instant and/or location. The importance of such a property has motivated a recent surge in interest in estimating the techno-economic value of EES in the power systems. For example, Sisternes et al. \cite{de2016value} argued that energy storage can potentially deliver value by increasing the cost-effective penetration of renewable energy in Texas (US), thus offering a new carbon-free alternative of operational flexibility. Shafiee et al. \cite{shafiee2016economic} assessed the benefits of energy storage as a price-maker to leverage certain arbitrage opportunities in the Alberta electricity market. Keck et al. \cite{keck2019impact} argued that large-scale ESS deployment with high renewable penetration could be economically feasible in Australia, hence facilitating the integration of renewable resources. Charles et al. \cite{charles2019sustainable} provided an investigation into the value of ESS to enhance the utility of small-scale generation assets for domestic solar home systems in rural Sub-Saharan Africa. Given a steadily decreasing trend of battery costs \cite{cole2019cost} and an increasing number of electrification policies promoting global sustainability \cite{doe}, ESS receives growing attention in the techno-economic portfolio of generation planning options toward a new energy future \cite{federal2016electric}.

Still, open questions remain at large about whether and how to invest and operate battery ESS economically while addressing their unique physical characteristics (e.g., nonlinear performance and degradation behavior) in the modeling of distributed generation portfolios. The analysis of distributed electricity system planning and operation is complex, not only because it requires a consideration of inter-temporospatial interactions but also because it is impacted by a variety of different tariff structures (if connected to the main grid). Significant efforts have thus been made to address the problem using either simulation-based \cite{mendes2011planning} or optimization-based methods \cite{fathima2015optimization}. In general, simulation-based analyses (e.g., HOMER \cite{homer}) provide fast and straightforward solutions that do not guarantee optimality; while optimization-based approaches (e.g., DER-CAM \cite{dercam}) usually generate the best feasible solutions at the cost of approximating nonlinear effects. System-level optimization analyses mainly focus on the planning and operation constraints of DERs under simplified grid and storage assumptions \cite{GARM2020}. Efforts to enhance the modeling details are mostly channeled toward adding more DER options or demand side management constraints \cite{TOORYAN2020}. Many of these existing tools usually do not consider a full year of power system operations in optimizing the capacity expansion decisions.

One of the key shortcomings in the literature is that existing DER planning models adopt simplified representations of the ESS with fixed physical characteristics. The behaviors of electrochemical batteries are extremely sensitive to operation \cite{bernardi1985general}. For example, the charge/discharge efficiencies and maximum powers are dependent on the battery's power input/output and state of charge \cite{sakti2017enhanced}. Many models fail to incorporate or consider these relationships and thus may generate misleading assessment of the economic benefits of the storage technologies. Moreover, the degradation cost cannot be neglected when operating the battery, and its effect must be considered in both planning and operation. Several studies have touched on modeling the performance and degradation of batteries. For instance, He et al. \cite{he2015optimal} developed a battery cycle life model to assess the optimal bids in day-ahead energy, spinning reserve, and regulation markets. D{\'\i}az et al. \cite{diaz2018maximum} considered battery aging in a dynamic programming approach to derive the optimal operations in certain grid applications. Sep{\'u}lveda-Mora and Hegedus \cite{sepulveda2021making} evaluated the impact that the battery degradation limit has on the cost-effectiveness of grid connected PV systems for commercial buildings. However, most existing literatures usually solve an operation problem that constrains the operational ﬂexibility of batteries. There are few studies incorporating detailed battery degradation into the investment planning problem. Furthermore, available studies only consider simplified assumptions, such as constraining the battery cycling to a degradation cap instead of optimizing the battery investment and operation based on its economic value \cite{cardoso2018battery}. 

To address the concerns outlined above, this study proposes an electricity generation planning model for distributed power systems with a particular focus on enhanced representation of ESS. The model selects the cost-minimal portfolio of generation capacity to reliably meet the electricity demand in a full year, accounting for detailed operational constraints as well as CO$_2$ emission goals. Several case studies are conducted for both off-grid and on-grid scenarios on a single-node distributed power system based on publicly available data. The major contributions of this study include: (1) incorporating a detailed battery ESS dynamic performances as well as cycle and calendar degradations model into a distributed power system planning problem; (2) exploring the impact of the battery model complexities on the optimal solutions; (3) investigating several decarbonization and ESS scenarios to analyze the optimal solutions toward low-carbon distributed power systems. Overall, it is expected that the developed framework will contribute to a more accurate assessment on the role of ESS benefits in future distributed power systems.

The rest of the paper is structured as follows: \textbf{Section 2} details the mathematical formulation of the model, including the distributed power system planning problem and several different ESS representations. \textbf{Section 3} introduces the case study with all the inputs and assumptions. We present the empirical results under different scenarios in \textbf{Section 4}, and the final remarks and conclusions in \textbf{Section 5}.

\begin{figure}[h]
\centering
\includegraphics[width=.425\linewidth , clip]{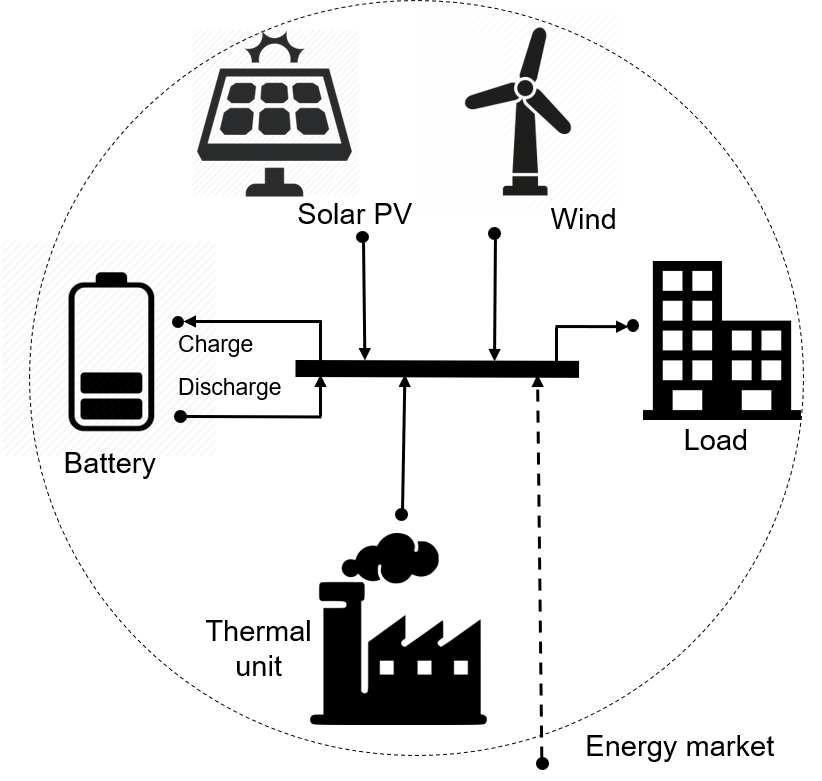}
\caption{A single-node distributed power system configuration with optional connection to the local grid.}
\end{figure}

\section{Mathematical formulation}

The main goal of this study is to evaluate the potential impacts of ESS in decarbonization of distributed power systems by enhanced ESS modeling approaches. It is important to note that we do not optimize the ESS in isolation, so investment and operation of different generation technologies and ESS are considered to serve a certain load. Therefore, without loss of generality, this study examines the behavior of a battery system combined with solar photovoltaic (PV), wind power, and a thermal power plant for our selected single-node case (see \textbf{Figure 1}). Note that a multi-node formulation can alter the role of ESS, particularly in large-scale power systems, but the focus of this study is on small-scale distributed power systems which can be represented by a single node. Also, aggregating variable renewable resources (PV and wind in this case) can decrease the variability in a large geographical area which leads to a lower storage requirement. However, since this study investigates a small geographical area, the aggregation should not affect the optimization results substantially. We start with a formulation of a basic battery model, which is widely used in the literature, to set a benchmark for comparison with the subsequently advanced battery models.

\subsection{Baseline model}

A conventional planning problem in the power grid is to minimize the system's total investment and operation costs. Here, the main focus is on the battery ESS formulation in a single-bus planning problem using mathematical programming as follows:

\begin{fleqn}
\begin{align}
& \text{min} && C_{\textrm{total}} = C_{\textrm{T}} + C_{\textrm{PV}} + C_{\textrm{W}} + C_{\textrm{B}} + C_{\textrm{LC}} + C_{\textrm{M}} \\
& \textrm{where} && C_{\textrm{T}} = I_{\textrm{T}} + OM_{\textrm{T}} = x \cdot \textrm{sc}_{\textrm{T}} \textrm{IF}_{\textrm{T}} \textrm{A}_{\textrm{T}} \textrm{TF} + x \cdot \textrm{sc}_{\textrm{T}} \textrm{FOM}_{\textrm{T}} \textrm{TF} + \sum_{t=1}^T g_{\textrm{T}}(t) \textrm{VOM}_{\textrm{T}}\\
&  && C_{\textrm{PV}} = I_{\textrm{PV}} + OM_{\textrm{PV}} = c_{\textrm{PV}} \textrm{IF}_{\textrm{PV}} \textrm{A}_{\textrm{PV}} \textrm{TF} + c_{\textrm{PV}} \textrm{FOM}_{\textrm{PV}} \textrm{TF} + \sum_{t=1}^T g_{\textrm{PV}}(t) \textrm{VOM}_{\textrm{PV}}\\
&  && C_{\textrm{W}} = I_{\textrm{W}} + OM_{\textrm{W}} = c_{\textrm{W}} \textrm{IF}_{\textrm{W}} \textrm{A}_{\textrm{W}} \textrm{TF} + c_{\textrm{W}} \textrm{FOM}_{\textrm{W}} \textrm{TF} + \sum_{t=1}^T g_{\textrm{W}}(t) \textrm{VOM}_{\textrm{W}}\\
&  && C_{\textrm{B}} = I_{\textrm{B}} + OM_{\textrm{B}} \nonumber \\
& &&\hspace{14pt}= (c_{\textrm{B}} \textrm{IP}_{\textrm{B}} + E_{\textrm{B}}\: \textrm{IE}_{\textrm{B}}) \textrm{A}_{\textrm{B}} \textrm{TF} + c_{\textrm{B}} \textrm{FOM}_{\textrm{B}} \textrm{TF} + \sum_{t=1}^T \left[ e_{\textrm{Bd}}(t) + e_{\textrm{Bc}}(t)\right]\textrm{VOM}_{\textrm{B}}\\
&  && C_{\textrm{LC}} = \sum_{t=1}^T c(t) \textrm{P}_\textrm{LC}\\
&  && C_{\textrm{M}} = \sum_{t=1}^T g_{\textrm{M}}(t) \textrm{P}_\textrm{M}(t)\\[12pt]
& \text{s.t.}
&& g_{\textrm{T}}(t) + g_{\textrm{PV}}(t) + g_{\textrm{W}}(t) + e_{\textrm{Bd}}(t) + g_{\textrm{M}}(t)  = \textrm{L}(t) + e_{\textrm{Bc}}(t) - c(t) \;\;\;\;\;\forall t\\
&&& g_{\textrm{T}}(t) \leq \textrm{AF}_{\textrm{T}}(t)x \cdot \textrm{sc}_{\textrm{T}}\;\;\;\;\;\forall t\\
&&& g_{\textrm{PV}}(t) \leq \textrm{AF}_{\textrm{PV}}(t)c_{\textrm{PV}}\;\;\;\;\;\forall t\\
&&& g_{\textrm{W}}(t) \leq \textrm{AF}_{\textrm{W}}(t)c_{\textrm{W}}\;\;\;\;\;\forall t\\
&&& e_{\textrm{Bc}}(t) = P_{\textrm{c}}(t)/\eta_\textrm{c}\;\;\;\;\;\forall t\\
&&& e_{\textrm{Bd}}(t) = P_{\textrm{d}}(t)\eta_\textrm{d}\;\;\;\;\;\forall t\\
&&& P_{\textrm{c}}(t) \leq \textrm{P}_{\textrm{c}}^{\textrm{max}}c_\textrm{B}w(t)\;\;\;\;\;\forall t
\end{align}
\end{fleqn}

\begin{fleqn}
\begin{align}
&\hspace{30pt}&& P_{\textrm{d}}(t) \leq \textrm{P}_{\textrm{d}}^{\textrm{max}}c_\textrm{B}[1-w(t)]\;\;\;\;\;\forall t\\
&&& E_{\textrm{B}} \textrm{SOC}_{\textrm{min}} \leq SOE(t) \leq E_{\textrm{B}} \textrm{SOC}_{\textrm{max}}\;\;\;\;\;\forall t\\
&&& SOE(1) = SOE(0) - P_{\textrm{d}}(1) + P_{\textrm{c}}(1)\\
&&& SOE(t) = SOE(t-1) - P_{\textrm{d}}(t) + P_{\textrm{c}}(t)\;\;\;\;\;\forall t \geq 2\\
&&& SOE(0) = E_{\textrm{B}} \textrm{SOC}_{\textrm{min}}\\
&&& (1-\lambda)SOE(0) \leq SOE(T) \leq (1+\lambda)SOE(0)\\
&&& g_{\textrm{D}}(t), g_{\textrm{M}}(t), g_{\textrm{PV}}(t), g_{\textrm{W}}(t), c(t), c_\textrm{B}, c_\textrm{PV}, c_\textrm{W}, E_\textrm{B}, e_{\textrm{Bc}}(t), e_{\textrm{Bd}}(t), P_{\textrm{c}}(t), P_{\textrm{d}}(t) \geq 0\;\;\;\;\;\forall t\\
&&& w(t) \in \{0, 1\}\;\;\;\;\;\forall t\\
&&& x \in \mathbb{Z}^{\geq 0} \;\;  
\end{align}
\end{fleqn}

The objective function (1) is defined as the system costs, which include the annualized investment and operation costs of each generation technology and ESS, load curtailment costs, and grid energy purchase costs (if grid connected). Equations (2)-(7) further detail the calculations of each cost element, where the thermal plant capacity is discrete while the capacity of PV, wind, and ESS are assumed to be continuous for simplicity. With Equation (5), the ESS power and energy capacities are optimized separately, leading to an optimal ESS duration requirement in the results. Note that participation in the electricity market will be enabled only when there is a connection to the main grid (i.e., on-grid mode). 

The general energy balance in the system is represented by constraint (8), whereas the dispatches of each generation technology are limited to their available capacities by constraints (9)-(11). Constraints (12) and (13) reflect the constant efficiency for charge and discharge at a given time step for the basic ESS model, where constant efficiency parameters $\eta_{\textrm{c}}$ and $\eta_{\textrm{d}}$ are used, respectively. Constraints (14) and (15) ensure that the battery's charge/discharge rate is lower than its limit (assumed constant here) and that it does not charge and discharge at the same time, enforced by the binary variable $w(t)$. Finally, the battery state of energy (SOE) is initialized and updated via constraints (16)-(20). Note that SOE (kWh) is the amount of energy in the battery and it is defined to distinguish from SOC (\% of the full capacity).

It is easy to observe that constraints (14) and (15) are nonlinear. Although some previous studies investigated under what conditions the binary variable is necessary to separate the charge and discharge based on a price signal \cite{li2015sufficient}, such conditions cannot be generalized well in this case. Therefore, in order to solve the model, these constraints need to be linearized. One simple heuristic is to replace constraint (14) with constraints (24) and (25), where M is a big number.

\begin{fleqn}[\parindent]
\begin{align}
& P_{\textrm{c}}(t) \leq \textrm{M} w(t)\\
& P_{\textrm{c}}(t) \leq \textrm{P}_{\textrm{c}}^{\textrm{max}}c_\textrm{B}
\end{align}
\end{fleqn}

Note that if $w(t)$ is zero, then constraint (24) ensures that $P_{\textrm{c}}(t)$ will be zero as well. On the other hand, if $w(t)$ is 1, this constraint is relaxed. Following the same logic, constraint (15) can be replaced by:

\begin{fleqn}[\parindent]
\begin{align}
& P_{\textrm{d}}(t) \leq \textrm{M} [1 - w(t)]\\
& P_{\textrm{d}}(t) \leq \textrm{P}_{\textrm{d}}^{\textrm{max}}c_\textrm{B}
\end{align}
\end{fleqn}

Selecting an appropriate value of M has been widely studied in optimization research~\cite{pineda}. For electricity system planning and operation, the criterion depends on the physical properties of the selected case. Based on all the considerations in our formulation, M is selected five times the peak load, which is large enough to assure sufficient investment in the battery power in low- or zero-emission scenarios and small enough to guarantee optimization stability.

Therefore, the baseline model of the planning problem is re-formulated to minimize objective (1) with respect to constraints (8)-(13) and (16)-(27). The resulting model is a mixed integer linear programming (MILP) problem.

\subsection{Advanced battery models}

In the basic formulation of battery model, the charge/discharge power limits and efficiencies are constant, which does not reflect the real performance of the battery. Also, the degradation due to cycling and time has not been considered in the basic model. Here, these features are incorporated into the ESS model through the following formulations to improve the representation of the battery dynamics in the planning problem.

\subsubsection{Dynamic efficiencies and power limits}
The battery's charge/discharge efficiencies are nonlinear functions of the battery input/output powers at different SOC levels. To capture these nonlinear behaviors, piece-wise linearization is considered based on our former original work \cite{sakti2017enhanced} and some improvements in the computational efficiency of the implementation \cite{jafari2019pes}. Binary variables are used to define the SOC dependency of the efficiency curves and power limits.

The maximum charge/discharge power limits, $\textrm{P}_{\textrm{c}}^{\textrm{max}}$ and $\textrm{P}_{\textrm{d}}^{\textrm{max}}$, are modeled as functions of the SOC, and the charge/discharge efficiencies, $\eta_\textrm{c}$ and $\eta_\textrm{d}$, are considered as functions of both charge/discharge power and SOC. To formulate this, the SOC percentage is partitioned as: $0 = \textrm{B}(1) < \textrm{B}(2) < ... < \textrm{B}(K) = 1$ with a binary activation function $u(t, k)$ where $u(t, k) = 1$ if $\beta(t) \in [\textrm{B}(k), \textrm{B}(k+1)]$ and $u(t, k) = 0$ otherwise. Detailed formulations are given as follows:

\begin{fleqn}[\parindent]
\begin{align}
& e_{\textrm{Bc}}(t) = P_{\textrm{c}}(t) + P_\textrm{c}^\textrm{loss}(t)\;\;\;\;\;\forall t\\[3pt]
& e_{\textrm{Bd}}(t) = P_{\textrm{d}}(t) - P_\textrm{d}^\textrm{loss}(t)\;\;\;\;\;\forall t\\[12pt]
& \beta (1) = \frac{SOE(0) + SOE(1)}{2\hat{\textrm{E}}_\textrm{B}}\\[6pt]
& \beta (t) = \frac{SOE(t - 1) + SOE(t)}{2\hat{\textrm{E}}_\textrm{B}}\;\;\;\;\;\forall t \geq 2\\[6pt]
& P_\textrm{c}(t) = \sum_{k=1}^K \sum_{j=1}^{J} v_\textrm{c}(k, j, t) \;\;\;\;\;\forall t\\
& P_\textrm{c}^\textrm{loss}(t) = \sum_{k=1}^K \sum_{j=1}^{J} v_\textrm{c}(k, j, t) \alpha_\textrm{c}(k, j) \;\;\;\;\;\forall t\\
& P_\textrm{d}(t) = \sum_{k=1}^K \sum_{i=1}^{I} v_\textrm{d}(k, i, t) \;\;\;\;\;\forall t\\
& P_\textrm{d}^\textrm{loss}(t) = \sum_{k=1}^K \sum_{i=1}^{I} v_\textrm{d}(k, i, t) \alpha_\textrm{d}(k, i) \;\;\;\;\;\forall t\\[3pt]
& \beta (t) \leq \sum_{k=1}^K u(t, k) \textrm{B}(k + 1)\;\;\;\;\;\forall t
\end{align}
\end{fleqn}

\begin{fleqn}[\parindent]
\begin{align}
& \beta (t) \geq \sum_{k=1}^K u(t, k) \textrm{B}(k)\;\;\;\;\;\forall t\\[3pt]
& v_\textrm{c}(k, j, t) \leq u(t, k) \textrm{PC}(k, j) \;\;\;\;\;\forall j, k, t\\[3pt]
& v_\textrm{d}(k, i, t) \leq u(t, k) \textrm{PD}(k, i) \;\;\;\;\;\forall i, k, t\\[3pt]
& \sum_{k=1}^K u(t, k) \leq 1\;\;\;\;\;\forall t\\[3pt]
& P_\textrm{c}^\textrm{loss}(t), P_\textrm{d}^\textrm{loss}(t), v_\textrm{c}(k, j, t), v_\textrm{d}(k, i, t) \geq 0\;\;\;\;\;\forall i, j, k, t\\[3pt]
& u(t, k) \in \{ 0, 1 \}\;\;\;\;\;\forall k, t
\end{align}
\end{fleqn}\vspace{-9pt}

In this improved model of the charge/discharge efficiency losses, constraints (12) and (13) in the baseline model are replaced with constraints (28) and (29), respectively. Constraints (30) and (31) calculate the average SOC in each time step using the battery's SOE values at the beginning and end of that time step and the total energy capacity of the battery. Note that, in order to avoid nonlinearity in constraints (30) and (31), an estimate of the battery energy capacity ($\hat{\textrm{E}}_\textrm{B}$) determined from the baseline model is used to calculate the SOC percentage. 

The charge/discharge powers are calculated by Equations (32) and (34) as the sum of linearized power curve pieces on the selected SOC curve. On the other hand, the charge/discharge losses are calculated by Equations (33) and (35), using the power curve pieces and their associated loss slopes. The corresponding input parameters can be obtained from ESS manufacturer data or experimental tests. Constraints (36) and (37) activate the selection of the associated SOC curve using the binary variable $u$. Finally, constraints (38) and (39) impose an upper limit on the linearized power curve pieces, and constraint (40) ensures that only one SOC curve is active at each time step.

Therefore, to model this advanced battery representation, the planning problem is formulated to minimize the objective (1) subject to constraints (8)-(11) and (16)-(42). This is still an MILP model.

\subsubsection{Battery degradations}

Incorporating the battery degradation into the model requires accounting for both calendar and cycling effects. Existing modeling approaches in many empirical studies address the calendar and cycling degradation independently to approximate the capacity loss by fitting polynomial, exponential, or other mathematical functions to the experimental data. This can be achieved by incorporating either one effect at each time step \cite{petit2016development}, by adding both aging components \cite{wang2016quantifying}, or by multiplying them \cite{baghdadi2016lithium}. However, from an electrochemical perspective, both aging phenomena occur simultaneously and capturing their exact details requires complicated experimental tests \cite{jafari2018deterministic}. Our recent study shows that different representations of battery degradation can substantially change the performance and revenue estimations for ESS operating in energy and capacity markets \cite{Jafari_Sakti20}.

In this paper, based on the method widely suggested in battery experimental studies \cite{SARASKETAZABALA2016839, WANG2014937, SCHMALSTIEG2014325}, it is considered that the total degradation can be calculated by adding the calendar and cycle degradations. A linear model is developed to capture both calendar and cycle degradations within the capacity expansion planning problem. Similar to \cite{Jafari_Sakti20}, it is assumed that the cycle degradation is a linear function of the number of equivalent full cycles which is calculated on a daily basis from the cycled energy of the ESS. The calendar degradation is assumed to be a linear function of time. The total battery degradation is the sum of the cycle and calendar degradation components (\textbf{Figure 2}). A weight index $p$ $\in[0,1]$ is defined to assign the contribution of each effect on the total degradation. Constraints (43)-(46) show how the degradation is formulated in the model.

\begin{fleqn}[\parindent]
\begin{align}
& E_\textrm{BD}(1) = E_\textrm{B}\\[9pt]
& E_\textrm{BD}(d) \leq E_\textrm{BD}(d-1) - \left[p\frac{\textrm{EOL}}{\textrm{N}_\textrm{cycle}} \sum_{t=24(d-2) + 1}^{24(d-1)} P_\textrm{c}(t) + (1-p)E_\textrm{B}\frac{\textrm{EOL}}{\textrm{l}_\textrm{B}}\right]\;\;\;\;\;\forall d \geq 2\\[9pt]
& E_\textrm{BD}(D) \leq E_\textrm{B} - \left[p\frac{\textrm{EOL}}{\textrm{N}_\textrm{cycle}} \sum_{t = 1}^{T}
P_\textrm{c}(t) + (1-p)E_\textrm{B}\frac{\textrm{TF}}{\textrm{l}_\textrm{B}} \textrm{EOL}\right]\\[9pt]
& E_{\textrm{B}} \textrm{SOC}_{\textrm{min}} \leq SOE(t) \leq E_{\textrm{BD}}(d) \textrm{SOC}_{\textrm{max}}\;\;\;\;\;\forall d, t
\end{align}
\end{fleqn}\vspace{-9pt}

\noindent where constraint (43) sets the battery capacity on day 1 equal to the installed capacity. Constraint (44) calculates the battery capacity on day $d$ considering the cycling and calendar degradation history in the previous days. The additional degradation constraint (45) ensures that the last day is also considered in the degradation calculations. Note that the cycle life N$_{\textrm{cycle}}$ and the calendar life l$_{\textrm{B}}$ in the constraint are input parameters of the optimization to avoid nonlinearity. With this model, there is a minimum degradation which is time-dependent and happens regardless of battery use, and any cycling of the battery will cause extra degradation. The battery capacity is updated every day and therefore the maximum SOE in constraint (16) is updated by (46) to account for the lost energy capacity. 

Finally, in order to incorporate the cost related to the battery degradation into the objective function, the term $E_{\textrm{B}} \textrm{IE}_{\textrm{B}}$ should be revised in the total battery cost $C_{\textrm{B}}$ in Equation (5). It is replaced with $E_{\textrm{BL}} \textrm{IE}_{\textrm{B}}$, the investment cost of the battery based on the lost capacity which comes from the degradation calculations as shown in (47)-(48). Note that the power component cost is not a function of degradation.

\begin{figure}[!t]
\centering
\includegraphics[width=.6\linewidth]{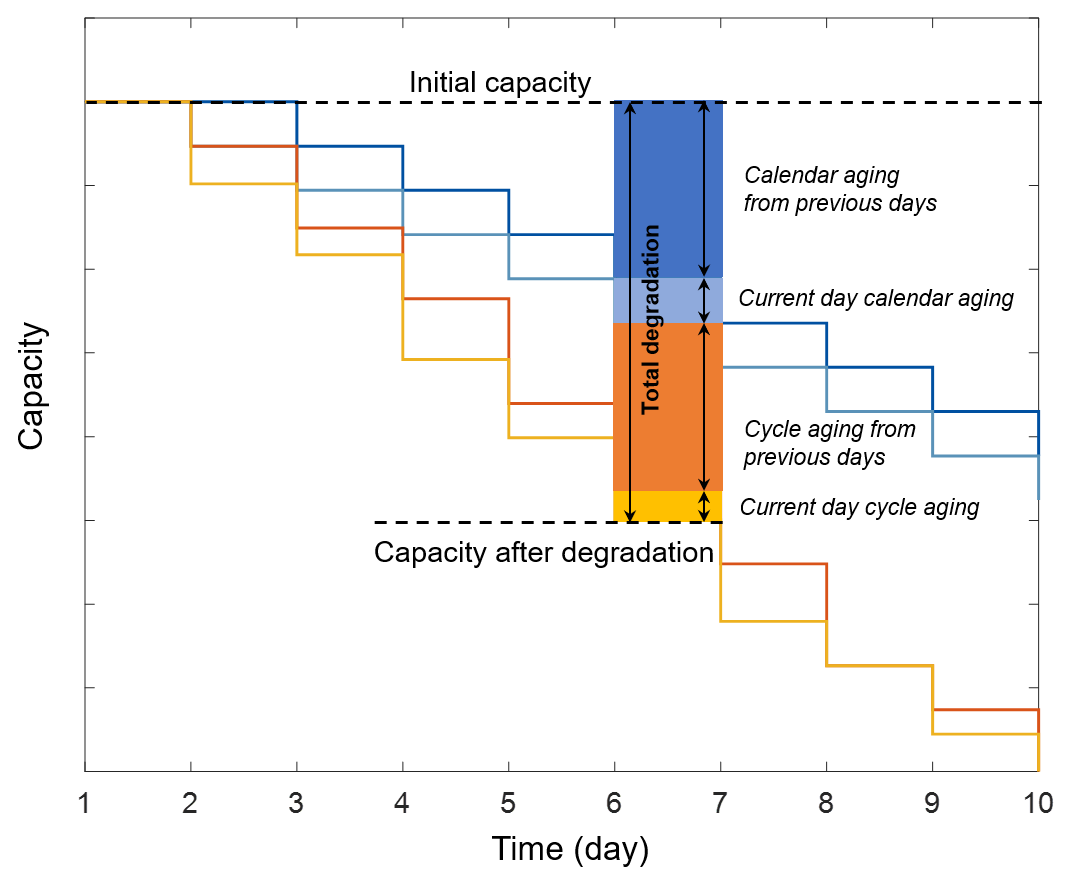}
\caption{Daily calendar and cycle degradation and the corresponding battery capacity evolution: daily calendar degradation is fixed (time-dependent); and daily cycle degradation is variable (use-dependent).}
\end{figure}

\begin{fleqn}[\parindent]
\begin{align}
& E_{\textrm{BL}} = \big[E_{\textrm{B}} - E_{\textrm{BD}}(D)\big]\: \frac{\textrm{l}_\textrm{B}}{\textrm{EOL}}\\[9pt]
&C_{\textrm{B}} = I_{\textrm{B}} + OM_{\textrm{B}} \nonumber \\
&\hspace{14pt}= (c_{\textrm{B}} \textrm{IP}_{\textrm{B}} \textrm{TF} + E_{\textrm{BL}}\: \textrm{IE}_{\textrm{B}}) \textrm{A}_{\textrm{B}} + c_{\textrm{B}} \textrm{FOM}_{\textrm{B}} \textrm{TF} + \sum_{t=1}^T \left[ e_{\textrm{Bd}}(t) + e_{\textrm{Bc}}(t)\right]\textrm{VOM}_{\textrm{B}}
\end{align}
\end{fleqn}

\subsection{Controlling carbon emissions}

The last important feature in the developed model is to add a constraint on the carbon emissions for studying different decarbonization scenarios. To do so, constraint (49) is added to the problem as follows: 

\begin{fleqn}[\parindent]
\begin{align}
& \sum_{t=1}^T g_{\textrm{T}}(t) \textrm{CO}_{\textrm{2 T}} + \sum_{t=1}^T g_{\textrm{M}}(t) \textrm{CO}_{\textrm{2 M}} \leq \gamma \: \textrm{CO}_{\textrm{2 total}}^{\textrm{unconstrained}}
\end{align}
\end{fleqn}\vspace{-6pt}

\noindent where, $\textrm{CO}_{\textrm{2 total}}^{\textrm{unconstrained}}$ is the unconstrained CO$_2$ emissions from a pure cost minimization case. $\gamma \in [0, 1]$ is a customized carbon emission percentage reduction compared to the unconstrained baseline. Note that the CO$_2$ emissions from the grid are calculated only when the system is connected to the local electricity network (i.e., on-grid mode). To summarize, the most advanced model is to minimize the objective function (1), with Equation (48) replacing (5), subject to constraints (8)-(11), (17)-(47), and (49). Equations (2)-(4), (6)-(7), and (48) are the expressions to define elements of the objective function. This model can be run with different levels of complexity for the battery representation. 

All these modeling formulations have been developed in Julia\footnote{Julia v0.6 was used in this study.} \cite{bezanson2017julia} and JuMP (Julia Mathematical Programming) \cite{dunning2017jump}. The commercial solver Gurobi \cite{gurobi} is used to solve the MILP optimization problem, with a customized MIP gap of 0.1\% in this study.

\section{Case study: inputs and assumptions}

A hypothetical site in Italy is considered with the electric load and day-ahead market information from ENTSO-E \cite{entso} and the renewable energy information from Renewables.ninja \cite{pfenninger2016long, staffell2016using} to investigate the decarbonization scenarios for a small-scale distributed power system with the developed ESS models. The market data was further calibrated\footnote{\label{ongrid}It is assumed that the grid cost consists of variable and fixed cost. The variable cost (i.e., energy cost) is charged according to the hourly prices of the day-ahead market; the fixed cost (i.e., capacity cost) is charged based on utility tariffs \cite{milan}. Our current case study assumes that the hourly grid supply is less than or equal to 100 kW. Then, the total fixed cost is a customer charge (\$84.87) plus a demand charge (\$14.11/kW for the demand exceeding 50 kW) for every month. } according to some utility tariffs in Milan, Italy \cite{milan}. The solar and wind resources are represented as hourly availability factors in the optimization. All the annual profiles are obtained from public historical or projected data. \textbf{Table 1} summarizes the input information for each case. It is important to note that the original profiles have been scaled so that the load remain at a distributed microgrid level with a peak hourly load of 573.3 kW. \added[id=mark]{We fix the electric demand profiles over time for the sake of a fair comparison and to keep the focus on energy storage in this study. Based on the projections \cite{entso}, the demand would increase 30\% from 2020 to 2050, which in turn should increase the investment and operation costs proportionally.} 

In order to account for potential price and policy changes, this study mainly experiments with two time scenarios (i.e., 2020 and 2050), two design strategies (i.e., on-grid and off-grid), and a transition case (i.e., 2020-2050) to investigate the resulting optimal generation portfolios. A small-scale open cycle gas turbine (OCGT) is considered to be the thermal plant in this study. Accordingly, the price information and input parameters for each technology are obtained based on \cite{cole2019cost, heather2018study, moro2018electricity, de2018technology, kairies2017battery}. \textbf{Table 2} presents an overview of the input parameters for different technologies in different time scenarios. Given the focus of the case study and the uncertainty of predicting future technology costs, three levels of the battery costs (i.e., low, medium, and high) are considered in 2050. The deployment of power limits and losses with piece-wise linear representations is shown in \textbf{Figure 3} for the advanced model in the case study. Three SOC levels (i.e., 20\%, 50\%, and 80\%) are considered using the data from \cite{sakti2017enhanced}. 

\begin{table}[!t]
\small
\begin{center}
\caption{Summary of the input data.}
\begin{tabular}{llllllllllll}
\toprule
                                  & \textbf{2020}  & \textbf{} & \textbf{} & \textbf{2030}  & \textbf{} & \textbf{} & \textbf{2040}  & \textbf{} & \textbf{} & \textbf{2050}    & \textbf{}        \\
                                  \cline{2-3} \cline{5-6} \cline{8-9} \cline{11-12} 
                                  & Max & Avg & \textbf{} & Max & Avg & \textbf{} & Max & Avg & \textbf{} & Max & Avg \\ \hline
Electric demand (kW)     & 573.3    & 371.2      &   & 573.3    & 371.2   &   & 573.3    & 371.2      &         & 573.3           & 371.2            \\
Day-ahead market (\$/kWh) & 0.26    & 0.10      &            & --            & --    &            & --            & -- &            & --            & --            \\
Solar availability      & 0.67      & 0.14      &   & 0.72      & 0.15      &  & 0.76      & 0.16      &          & 0.81           & 0.17             \\
Wind availability        & 1       & 0.20       &   & 1       & 0.22       &  & 1       & 0.24       &          & 1               & 0.25             \\
\bottomrule
\end{tabular}
\end{center}\vspace{12pt}
Note: The electric load and day-ahead market data are obtained from ENTSO-E historical data \cite{entso}, and the solar and wind availabilities are obtained from Renewables.ninja \cite{pfenninger2016long, staffell2016using}. The availability factor profiles are scaled up from 2020 to 2050 to reflect technology progress, keeping the hourly variations. Due to the high uncertainty of predicting the market price, we do not consider on-grid mode for future scenarios in the case study.
\end{table}

\begin{figure}[!b]
\centering
\includegraphics[width=.55\linewidth]{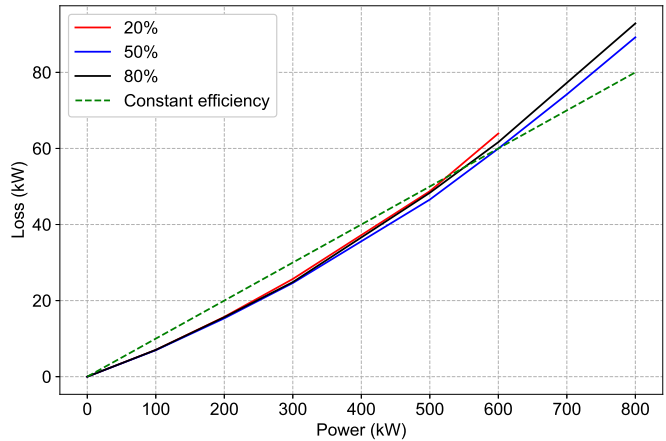}
\caption{Piece-wise linearized charge/discharge losses with respect to input/output power at different SOC levels \cite{sakti2017enhanced}.}
\end{figure}

\begin{table}[!t]
\small
\caption{Summary of the input parameters for each component.}
\begin{center}
\begin{tabular}{lllll}
\toprule
\textbf{Parameter}                     & \textbf{2020  }  & \textbf{2030  }  & \textbf{2040  }   & \textbf{2050  } \\ \hline
\rule{0pt}{3ex}\textit{Grid}\vspace{2pt}                 &               &  &           \\
\hspace{9pt}Load curtailment (\$/kWh)    & 13            & 13   & 13     & 13       \\
\hspace{9pt}Discount rate (\%)                 & 10           & 10 & 10  & 10           \\
\hspace{9pt}Unit CO$_2$ emissions (kg/kWh)    & 0.36          & -- & --  & --         \vspace{3pt} \\
\textit{Thermal plant}\vspace{3pt}                   &               &  &               \\
\hspace{9pt}Unit capacity (kW)            & 50            & 50 & 50  & 50            \\
\hspace{9pt}Investment cost (\$/kW)            & 1032.9        & 1013.1 & 1008.7 & 1004.3        \\
\hspace{9pt}Life time (year)              & 20            & 20 & 20  & 20            \\
\hspace{9pt}FOM cost (\$/kW year)              & 25.85         & 22 & 20.7   & 19.36         \\
\hspace{9pt}VOM cost (\$/kWh)                 & 0.0008         & 0.0008 & 0.0008 & 0.0008       \\
\hspace{9pt}Fuel cost (\$/kWh)           & 0.0595        & 0.0813 & 0.103  & 0.1246        \\
\hspace{9pt}Unit CO$_2$ emissions (kg/kWh)    & 0.52          & 0.51 & 0.5  & 0.49         \vspace{3pt} \\
\textit{PV}\vspace{3pt}                   &               &  &               \\
\hspace{9pt}Investment (\$/kW)            & 781           & 729.3 & 570.9   & 499.4         \\
\hspace{9pt}Life time (year)              & 25            &25 &25  & 25            \\
\hspace{9pt}FOM cost (\$/kW year)              & 13.86         & 11.88 & 11  & 10.12         \\
\hspace{9pt}VOM cost (\$/kWh)                 & 0             &0 &0  & 0            \vspace{3pt} \\
\textit{Wind}\vspace{3pt}                 &               &  &               \\
\hspace{9pt}Investment cost (\$/kW)            & 1424.5        & 1277.1 & 1111   & 1037.3        \\
\hspace{9pt}Life time (year)              & 25            & 25 &25 & 25            \\
\hspace{9pt}FOM cost (\$/kW year)              & 15.4          & 14.9 & 14.3  & 13.2          \\
\hspace{9pt}VOM cost (\$/kWh)                 & 0.0002        &0.0002 & 0.0002  & 0.0002       \vspace{3pt} \\
\textit{Battery}\vspace{3pt}              &               &  &               \\
\hspace{9pt}Max charging power (\%)         & 100             & 100 & 100 & 100             \\
\hspace{9pt}Max discharging power (\%)      & 100              & 100 & 100  & 100             \\
\hspace{9pt}Wrap-up tolerance              & 0.1           & 0.1 & 0.1  & 0.1           \\
\hspace{9pt}SOC upper limit               & 0.9           & 0.9 & 0.9  & 0.9           \\
\hspace{9pt}SOC lower limit               & 0.1           & 0.1 & 0.1  & 0.1           \\
\hspace{9pt}Power investment (\$/kW)      & 510           &370 & 330  & 140(L) / 280(M) / 470(H)          \\
\hspace{9pt}Energy investment (\$/kWh) & 150           & 120 & 100 & 40(L) / 80(M) / 145(H)           \\
\hspace{9pt}Cycle life                    & 3500          &4750 & 6000  & 7250          \\
\hspace{9pt}Lifetime (year)               & 13.6          &18.4 & 23.3  & 28.2          \\
\hspace{9pt}FOM cost (\$/kW year)              & 8             &7.5 & 6.5  & 2.5(L) / 6.5(M) / 8(H)         \\
\hspace{9pt}VOM cost (\$/kWh)                 & 0.0024        &0.002 &0.0013  & 0.0005(L) / 0.0013(M) / 0.0024(H)        \\
\hspace{9pt}Charge efficiency       & 0.9          & 0.9 & 0.9  & 0.9          \\
\hspace{9pt}Discharge efficiency     & 0.9          & 0.9 & 0.9  & 0.9          \\
\hspace{9pt}End-of-life criterion         & 70\%           & 70\% & 70\%  & 70\%           \\
\bottomrule
\end{tabular}
\end{center}\vspace{12pt}
Note: The load curtailment is considered from Ref. \cite{heather2018study}. The grid carbon emission is calculated according to Ref. \cite{moro2018electricity}. The information of the thermal plant (OCGT), PV, and wind is obtained from Ref. \cite{de2018technology}. The information of the battery is obtained from Refs. \cite{cole2019cost, kairies2017battery}. L, M, and H represent low, medium, and high projected prices, respectively. \\
\end{table}

\section{Result and discussion}

With the selected case study and input data, different scenarios are tested using the developed battery model. In particular, we are interested in the optimal generation portfolio, operational patterns, system costs, and carbon emissions. 

\subsection{Impact of ESS modeling complexity}

The first exercise is to examine the impact of different ESS modeling approaches on the optimal solutions. In our previous studies \cite{sakti2017enhanced, jafari2019pes}, the dynamic efficiencies and power limits were applied to the energy arbitrage problem. Here, the potential benefits of incorporating the dynamic efficiencies and power limits as well as the battery degradations into the generation planning problem are investigated. The differences these models make on the optimal expansion results and the corresponding computational burden are explored. Four versions of the model are configured with different levels of complexity in the ESS representation for an off-grid case without CO$_2$ emission constraint using 2050 cost projections. Note that, due to excessive computational time, only the first quarter of 2050 data is used here and in \textbf{Subsection 4.2}. However, for the rest parts, the results are obtained using the data over a full year (i.e., 8760 hours). \vspace{12pt}

\begin{itemize}
\item \textbf{Model 1}: Baseline model

\item \textbf{Model 2}: Baseline model + dynamic efficiencies and power limits

\item \textbf{Model 3}: Baseline model + degradation model

\item \textbf{Model 4}: Baseline model + dynamic efficiencies and power limits + degradation model
\end{itemize}\vspace{12pt}

In general, increasing the level of ESS modeling complexity will improve the accuracy of the results, and therefore, Model 4 should produce the most reliable solutions. However, additional modeling complexity will also increase the computational cost. The results for different models are compared to illustrate the trade-off between modeling accuracy and computational efficiency. As shown in \textbf{Figure 4} and \textbf{Table 3}, the resulting portfolios of Models 1 and 2 are very close to each other and those of Models 3 and 4 are quite similar as well. However, there is a substantial difference between Models 1-2 and Models 3-4. The average difference in the installed capacities of different technologies between Models 1 and 2 or Models 3 and 4 is 2\%, while it goes up to 65\% between Models 1-2 and 3-4. In particular, the installed capacities of the thermal plant (kW) and battery (kWh) change significantly in the latter comparison (+16\% and -170\%, respectively). This change leads to 17\% increase in the thermal plant's total (fixed + variable) costs and 47\% decrease in the battery's total costs, while the total system costs change less than 3\%. Due to the increased role of thermal capacity, the CO$_2$ emissions deviate 16\% between the optimal solutions for Models 1-2 and 3-4. These results indicate that, under the cost assumptions in this study, the impact of battery degradation seems more significant than the effect of dynamic efficiencies and power limits. Contrary to our previous findings on the energy arbitrage application \cite{sakti2017enhanced, jafari2019pes}, the impact of dynamic efficiencies and power limits seems considerably less significant for the generation portfolio planning problem. However, incorporating the battery degradation substantially alters the optimal generation portfolio, leading to a more accurate estimation of the renewable penetrations and CO$_2$ emissions.

\begin{figure}[!b]
\makebox[\textwidth][c]{
\centering
\includegraphics[width=1.0\linewidth]{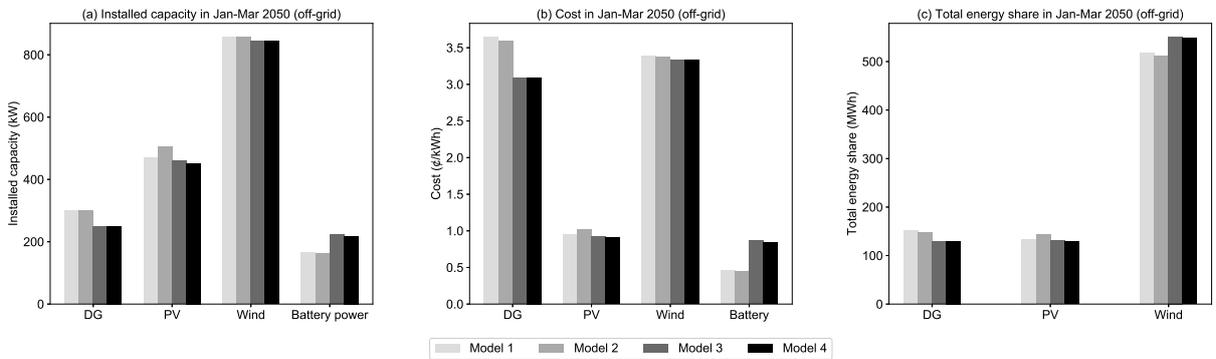}}
\caption{Comparison of optimal solutions of Models 1-4 for off-grid case in the first quarter of 2050.}
\end{figure}

\begin{table}[!h]
\small
\begin{center}
\caption{Detailed investment portfolios of Models 1-4 for off-grid case in the first quarter of 2050.}
\begin{tabular}{lllllllll}
\toprule
\textbf{Variable}          & \textbf{Unit} & \textbf{Model 1} & \textbf{} & \textbf{Model 2} & \textbf{} & \textbf{Model 3} & \textbf{} & \textbf{Model 4} \\ \hline
\rule{0pt}{3ex}\textit{Installed capacity}\vspace{2pt} &               &                  &           &                  &           &                  &           &                  \\
\hspace{9pt}Thermal plant                          & kW            & 300              &           & 300              &           & 250              &           & 250              \\
\hspace{9pt}PV                          & kW            & 472              &           & 505              &           & 459              &           & 452              \\
\hspace{9pt}Wind                        & kW            & 858              &           & 857              &           & 845              &           & 844             \\
\hspace{9pt}Battery power               & kW            & 167              &           & 162              &           & 223              &           & 218               \\
\hspace{9pt}Battery energy              & kWh           & 995               &           & 953             &           & 3,645             &           & 3,558             \vspace{3pt} \\
\textit{Cost \& emission}\vspace{3pt}   &               &                  &           &                  &           &                  &           &                  \\
\hspace{9pt}Total                       & \$            & 67,506            &           & 67,387            &           & 65,615            &           & 65,367            \\
\hspace{9pt}Thermal plant                          & \$            & 29,187            &           & 28,701            &           & 24,668            &           & 24,736             \\
\hspace{9pt}PV                          & \$            & 7,575             &           & 8,107             &           & 7,378             &           & 7,259             \\
\hspace{9pt}Wind                        & \$            & 27,072            &           & 27,032            &           & 26,662            &           & 26,624            \\
\hspace{9pt}Battery                     & \$            & 3,673             &           & 3,544             &           & 6,908             &           & 6,748              \\
\hspace{9pt}Load curtailment            & \$            & 0                 &           & 4.26              &           & 0                 &           & 0                \\
\hspace{9pt}CO$_2$ emissions            & kg             & 74,269            &           & 72,371            &           & 63,238            &           & 63,506  \vspace{3pt} \\
\textit{Energy share}\vspace{3pt}       &               &                  &           &                  &           &                  &           &                  \\
\hspace{9pt}Thermal plant                          & \%           & 19           &           & 18.5           &           & 16.2           &           & 16.2            \\
\hspace{9pt}PV                          & \%           & 16.8           &           & 17.9            &          & 16.3           &         & 16.1            \\
\hspace{9pt}Wind                        & \%           & 64.9           &           & 64.0           &           & 68.9           &           & 68.7           \\
\hspace{9pt}Battery charge              & \%           & -3.3            &           & -3.2           &           & -7.5            &           & -7.1            \\
\hspace{9pt}Battery discharge           & \%           & 2.7            &           & 2.8            &           & 6.0            &           & 6.1             \\
\hspace{9pt}Load curtailment            & \%           & 0                 &           & 0.00004              &           & 0                 &           & 0                \\
\bottomrule
\end{tabular}
\end{center}\vspace{12pt}
Note: The peak electricity load in the first quarter of 2050 is 525 kW. The total electricity load in the first quarter of 2050 is 799.6 MWh.
\end{table}

On the other hand, the numbers of decision variables of Models 1-4 are summarized in \textbf{Table 4}. The dynamic efficiencies and power limits model incurs much more variables, especially binary variables, which inevitably make for a more numerically intense problem for the MILP solver -- a very high computational cost is being paid for an insignificant gain in the performance.

Considering the optimal solutions and the computational costs of each model, it is concluded that the dynamic efficiencies and power limits of the battery seem not very critical for generation planning optimization. However, the battery degradation presents a significant impact and should be incorporated in the investment decision-making process. Therefore, Model 1 is too simplistic; Model 2 increases the computational cost without adding significant improvements to the results; Model 4 gives the most accurate solutions, but it is computationally expensive; Model 3 improves the accuracy of the results substantially with moderate computational burden. For the rest of our analyses, Model 3 (baseline model + degradation model) will be used.

\begin{table}[!b]
\small
\begin{center}
\caption{Number of decision variables in Models 1-4 of the case study.}
\begin{tabular}{lllll}
\toprule
\textbf{Variable} & \textbf{Model 1} & \textbf{Model 2}    & \textbf{Model 3} & \textbf{Model 4}        \\ \hline
Continuous               & $9T + 5$           & $(12 + IK + JK)T + 5$ & $9T + D + 5$       & $(12 + IK + JK)T + D + 5$ \\
Integer                  & 1                & 1                   & 1                & 1                       \\
Binary                   & $T$                & $(K+1)T$              & $T$                & $(K+1)T$                  \\
\bottomrule
\end{tabular}
\end{center}\vspace{12pt}
Note: $T$ is the number of hours, $D$ is the number of days, $K$ is the number of SOC levels, $J$ is the maximum number of pieces in the charge curves, and $I$ is the maximum number of pieces in the discharge curves.
\end{table}

\subsection{Impact of cycle and calendar degradation dominance}

As discussed in \textbf{Subsection 2.2.2}, battery degradation is calculated as a weighted sum of cycle and calendar degradation, based on the battery experimental studies. However, there is no straightforward guideline on the fraction of each factor in the total degradation effects, as represented in constraint (44). A sensitivity analysis is therefore performed for different values of $p$ to show its impact on the results. \textbf{Table 5} shows the optimal results for the selected fractions of cycle and calendar degradation in the total degradation calculation. A smaller weight for cycle degradation ($p=0.25$) means a larger constant cost of the battery (i.e., time-dependent calendar degradation), which leads to lower degree of freedom in optimizing its costs by cycling and therefore the installed battery capacity is smaller. On the other hand, when the cycle degradation has larger impact ($p=0.75$), the battery cost is mainly controlled by the cycling and, therefore, the optimal solution is to invest in larger battery capacity, especially in terms of energy (kWh). This leads to a reduction in the number of cycles, which in turn minimizes the total cost. Middle values for $p$ balances the impact of constant inevitable calendar degradation and variable cycling-related costs. In the following presented results, $p$ is set to 0.5. 

\begin{table}[!t]
\small
\begin{center}
\caption{Detailed investment portfolios of Model 3 with different cycle degradation weights for off-grid case in the first quarter of 2050.}
\begin{tabular}{lllllll}
\toprule
\textbf{Variable}          & \textbf{Unit} & \textbf{\textit{p} = 0.25} & \textbf{} & \textbf{\textit{p} = 0.5} & \textbf{} & \textbf{\textit{p} = 0.75} \\ \hline
\rule{0pt}{3ex}\textit{Installed capacity}\vspace{2pt} &               &                  &           &                  &           &                                   \\
\hspace{9pt}Thermal plant                          & kW            & 250              &           & 250              &           & 150                         \\
\hspace{9pt}PV                          & kW            & 633              &           & 459              &           & 581                         \\
\hspace{9pt}Wind                        & kW            & 840              &           & 845              &           & 954                       \\
\hspace{9pt}Battery power               & kW            & 224              &           & 223              &           & 326                          \\
\hspace{9pt}Battery energy              & kWh           & 2,546               &           & 3,645             &           & 12,049                        \vspace{3pt} \\
\textit{Cost}\vspace{3pt}   &               &                  &           &                  &           &                             \\
\hspace{9pt}Total                       & \$            & 66,714            &           & 65,615            &           & 62,316                     \\
\hspace{9pt}Battery                     & \$            & 6,595             &           & 6,908             &           & 12,350                         \\
\bottomrule
\end{tabular}
\end{center}\vspace{12pt}
Note: The peak electricity load in the first quarter of 2050 is 525 kW. The total electricity load in the first quarter of 2050 is 799.6 MWh.
\end{table}

\subsection{Decarbonization for off-grid case in 2020}

Now, several off-grid cases are presented for 2020 to illustrate the potential impact of energy storage on the generation portfolio at different decarbonization levels. Using Model 3 with the selected input data and assumptions over a full year, the cost minimization tasks are conducted for a range of CO$_2$ emission constraints. 

\textbf{Table 6} and \textbf{Figure 5} present the results of these scenarios. Note that the CO$_2$ constraints in these results are 100\% (i.e., no constraint on the CO$_2$ emissions), 75\%, 50\%, 25\%, and 0\% of the no-constraint value.
It can be seen that the thermal plant has the major portion of the installed capacity and the energy supply in the case with no constraint on the CO$_2$ emissions (i.e., 100\%), due mainly to its high availability factor and relatively low cost per kWh. On the other hand, as less CO$_2$ emissions are required, the renewable energy resources (i.e., PV and wind) are increasingly favored.  Consequently, the total system costs gradually increase with a more stringent carbon constraint.

\begin{figure}[!b]
\makebox[\textwidth][c]{
\centering
\includegraphics[width=1.0\linewidth]{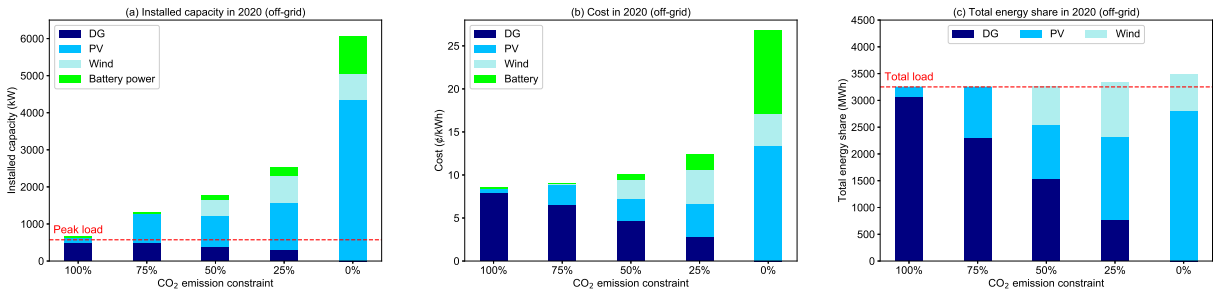}}
\caption{Optimal system configurations and costs for off-grid case in 2020 at different decarbonization levels.}
\end{figure}

\begin{table}[!t]
\small
\begin{center}
\caption{Detailed investment portfolios for off-grid case in 2020 at different decarbonization levels.}
\begin{tabular}{lllllllllll}
\toprule
\textbf{Variable}          & \textbf{Unit} & \textbf{100\%} & \textbf{} & \textbf{75\%} & \textbf{} & \textbf{50\%} & \textbf{} & \textbf{25\%} & \textbf{} & \textbf{0\%}\\ \hline
\rule{0pt}{3ex}\textit{Installed capacity}\vspace{2pt} &               &                &           &               &           &               &           &  \\
\hspace{9pt}Thermal plant                          & kW            & 500            &           & 500           &           & 400           &           & 300     &           & 0    \\
\hspace{9pt}PV                          & kW            & 143            &           & 774        &              & 819          &            & 1,277    &           & 4,353    \\
\hspace{9pt}Wind                        & kW            & 0              &           & 0             &           & 433           &           & 728  &           & 712   \\
\hspace{9pt}Battery power               & kW            & 31             &           & 27            &           & 114             &      & 231  &           & 1,007     \\
\hspace{9pt}Battery energy              & kWh           & 157            &           & 88           &       & 762        &           & 2,310  &    & 18,040        \vspace{3pt}  \\
\textit{Cost \& emission}\vspace{3pt}   &               &                &           &               &           &               &           &               \\
\hspace{9pt}Total                       & \$           & 277,782         &       & 293,828        &       & 329,735        &           & 404,132  &           & 871,462      \\
\hspace{9pt}Thermal plant                          & \$           & 258,945         &           & 212,605        &      & 151,549        &           & 90,492 &           & 0         \\
\hspace{9pt}PV                          & \$           & 14,308          &          & 77,276         &         & 81,835       &          & 127,655 &           & 434,861        \\
\hspace{9pt}Wind                        & \$           & 0              &           & 0             &           & 74,820         &      & 125,701 &           & 122,730        \\
\hspace{9pt}Battery                     & \$           & 4,131       &           & 3,337          &           & 20,657        &           & 60,173 &         & 313,871         \\
\hspace{9pt}Load curtailment            & \$           & 398            &           & 611           &           & 873           &           & 113  &           & 0           \\
\hspace{9pt}CO$_2$ emissions            & kg            & 1,589,352        &           & 1,192,014       &           & 794,676        &           & 397,338  &           & 0      \vspace{3pt}  \\
\textit{Energy share}\vspace{3pt}       &               &                &           &               &           &               &           &               \\
\hspace{9pt}Thermal plant               & \%           & 94.5    &           & 70.9       &       & 47.2       &           & 23.6 &           & 0        \\
\hspace{9pt}PV                          & \%          & 5.5          &      & 29.2        &          & 31.4        &      & 47.9 &           & 86.1      \\
\hspace{9pt}Wind                        & \%           & 0         &            & 0             &           & 21.9       &           & 31.1 &          & 20.9       \\
\hspace{9pt}Battery charge              & \%           & -0.06      &           & -0.2         &           & -3.0             &     & -13.9 &           & -37.2      \\
\hspace{9pt}Battery discharge           & \%           & 0.05    &           & 0.17         &           & 2.4             &       & 11.2 &           & 30.1        \\
\hspace{9pt}Load curtailment            & \%           & 0.0009             &           & 0.0014           &           & 0.0021            &           & 0.0003 &           & 0            \\
\textit{Curtailment}\vspace{3pt}       &               &                &           &               &           &               &           &               \\
\hspace{9pt}PV                          & \%          & 0          &      & 2.1        &          & 0.6        &      & 2.8 &           & 48.7      \\
\hspace{9pt}Wind                        & \%           & --         &            & --             &           & 6.0       &           & 20.6 &          & 45.4       \\

\bottomrule
\end{tabular}
\end{center}\vspace{12pt}
Note: The peak electricity load in 2020 is 573 kW. The total electricity load energy in 2020 is 3,252 MWh.
\end{table}

When larger PV and wind systems are added under the stricter CO$_2$ constraints, battery investments start occurring to reduce renewables curtailment. Note that, because of energy losses in the battery, the total energy output is slightly higher than the total load for the cases with ESS (see \textbf{Figure 5(c)}). However, considering the relatively high costs of batteries in 2020, ESS installations are relatively limited unless the  CO$_2$ emission constraint is set to 50\% or stricter.

If the unconstrained scenario is considered as the benchmark, stricter CO$_2$ constraints will increase the total costs by 5.8\%, 18.7\%, 45.5\%, and 214\% for the 75\%, 50\%, 25\%, and 0\% CO$_2$ emission constraints, respectively. Achieving a zero-emission system is substantially more expensive that 75\% decarbonization. The unit cost of electricity increases from 9 \textcent/kWh to 12 \textcent/kWh for 75\% CO$_2$ emission reduction. However, forcing zero-emission in the optimal solution increases the energy cost to 27 \textcent/kWh. It is observed that, although the thermal plant capacity remains the same between the 100\% and 75\% cases, its energy share drops substantially from 94.5\% to 70.9\%, owing to the limit on CO$_2$ emissions. Also, PV plays a more important role than wind in these results, especially for the zero-emission case. Note that the CO$_2$ intensity of the optimal solution without any carbon constraint is 489 g/kWh.  

\begin{figure}[!t]
\centering
\includegraphics[width=.84\linewidth]{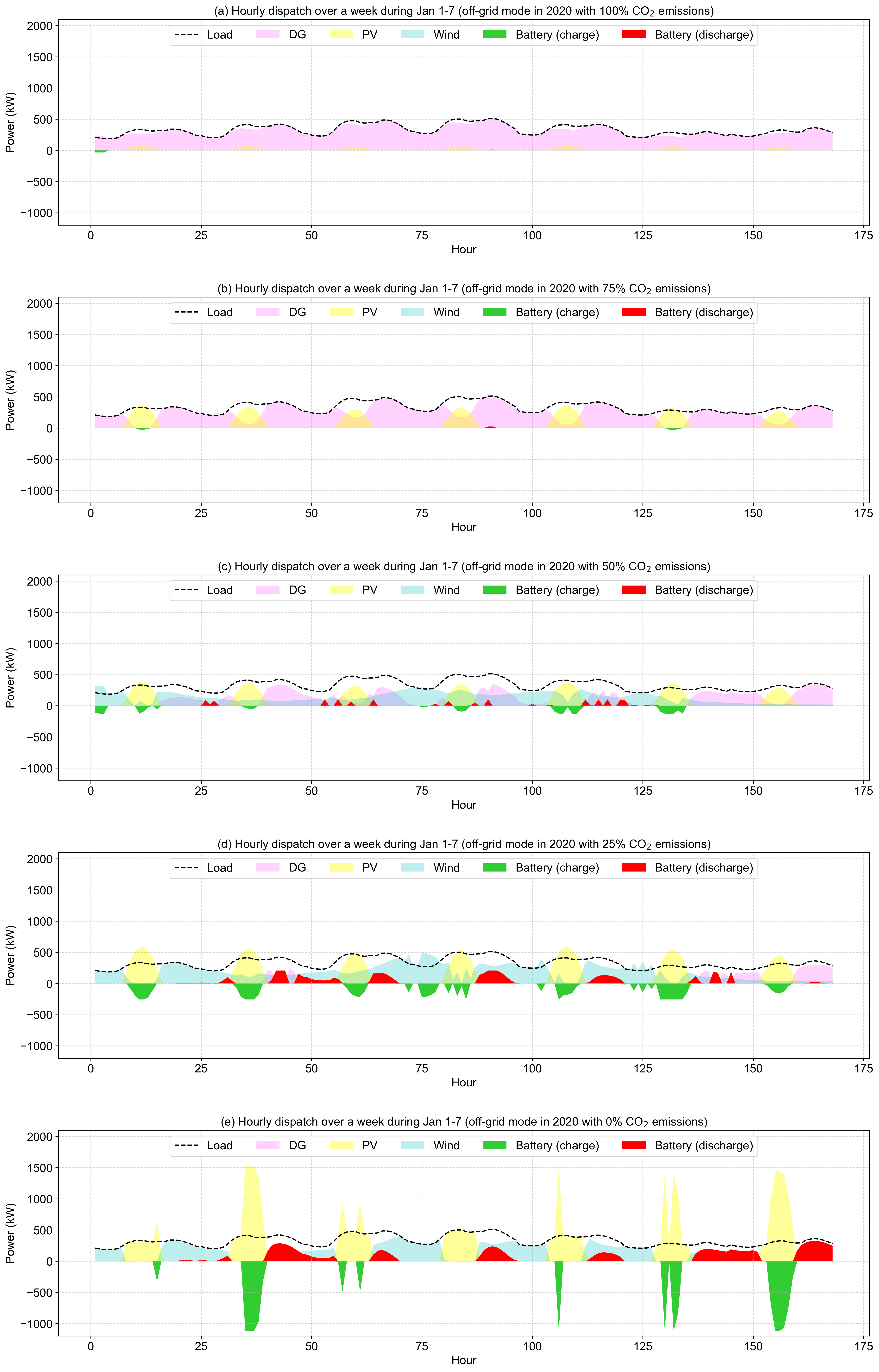}
\caption{Hourly electricity dispatch during January 1-7 for off-grid case in 2020 at different decarbonization levels.}
\end{figure}

A sample optimal dispatch of electricity from each case over January 1-7 (typical winter week) is depicted in \textbf{Figure 6}. An obvious pattern is that the CO$_2$ constraint will promote the use of PV, wind, and batteries so that less thermal generation is dispatched in order to reduce the CO$_2$ emissions. In this scenario, the battery is used for storing extra energy output mainly from PV and occasionally from wind. Dispatch result for the zero-emission system shows that the PV and wind outputs are very frequently curtailed, leading to 49\% and 45\% curtailment rates in this case. Since this is a zero-emission scenario, the system is 100\% renewable and the optimal solution does not invest in the thermal power plant. Therefore, it leads to higher PV and wind capacities as investment in renewables even with curtailment of the excessive generation is less expensive than investment in extra storage. Another issue that can arise with 100\% renewable system is stability. However, this case study is a small-scale distributed power system with the possibility of connecting to a larger grid with stability measures. Therefore, for simplicity, the stability can be provided by the grid in on-grid scenarios and also through virtual inertia from the battery and wind turbines in off-grid scenarios \cite{wogrin2020}.

\subsection{Impact of local grid connection in 2020}

\begin{figure}[!t]
\makebox[\textwidth][c]{
\centering
\includegraphics[width=1.0\linewidth]{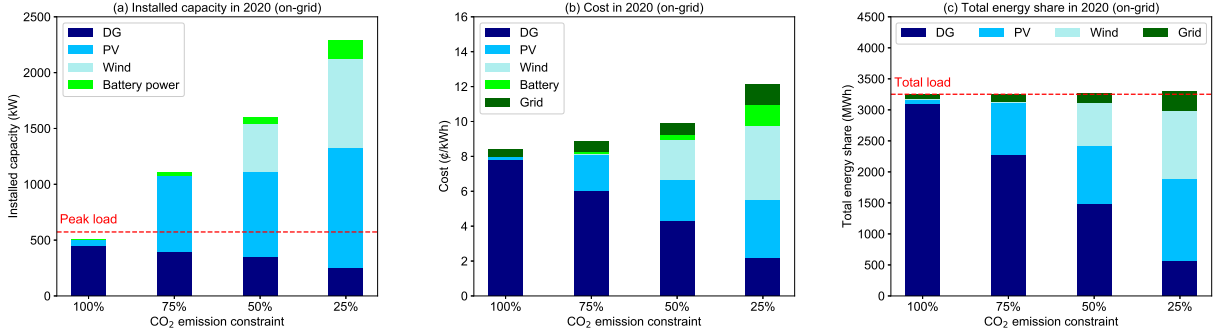}}
\caption{Optimal system configurations and costs for on-grid case in 2020 at different decarbonization levels. The grid capacity is constrained to 100 kW.}
\end{figure}

\begin{table}[!b]
\small
\begin{center}
\caption{Detailed investment portfolios for on-grid case in 2020 at different decarbonization levels.}
\makebox[\textwidth][c]{
\begin{tabular}{lllllllll}
\toprule
\textbf{Variable}          & \textbf{Unit} & \textbf{100\%} & \textbf{} & \textbf{75\%} & \textbf{} & \textbf{50\%} & \textbf{} & \textbf{25\%} \\ \hline
 \rule{0pt}{3ex}\textit{Installed capacity}\vspace{2pt} &                &           &               &           &               &           &               \\
 \hspace{9pt}Thermal plant                          & kW            & 450            &           & 400           &           & 350           &           & 250           \\
 \hspace{9pt}PV                          & kW            & 57             &           & 682           &           & 763           &           & 1,079          \\
 \hspace{9pt}Wind                        & kW            & 0              &           & 0             &           & 428          &            & 796           \\
 \hspace{9pt}Battery power               & kW            & 0              &           & 28            &           & 62             &          & 163           \\
 \hspace{9pt}Battery energy              & kWh           & 0              &           & 87            &           & 307           &           & 1,468    \vspace{3pt} \\
 \textit{Cost \& emission}\vspace{3pt}   &               &                &           &               &           &               &           &               \\
 \hspace{9pt}Total                       & \$            & 273,947        &           & 288,749       &           & 321,975       &           & 394,304       \\
 \hspace{9pt}Thermal plant                          & \$            & 253,384        &           & 196,318       &           & 140,646       &           & 71,484        \\
 \hspace{9pt}PV                          & \$            & 5,719          &           & 68,147        &           & 76,243        &           & 107,812       \\
 \hspace{9pt}Wind                        & \$            & 0              &           & 0             &           & 73,818        &           & 137,410       \\
 \hspace{9pt}Battery                     & \$            & 0              &           & 3,363         &           & 9,676         &           & 39,636        \\
 \hspace{9pt}Load curtailment            & \$            & 92             &           & 610           &           & 574           &           & 371           \\
 \hspace{9pt}Grid                        & \$            & 14,752         &           & 20,309        &           & 21,017        &           & 37,863        \\
 \hspace{9pt}CO$_2$ emissions             & kg            & 1,632,542      &           & 1,224,406     &           & 816,271       &           & 408,136      \vspace{3pt} \\
 \textit{Energy share}\vspace{3pt}       &               &                &           &               &           &               &           &               \\
  \hspace{9pt}Thermal plant               & \%          & 95.4      &           & 70.1     &           & 45.4     &           & 17.7       \\
 \hspace{9pt}PV                          & \%           & 2.2         &           & 26.0       &           & 29.1       &           & 40.7     \\
 \hspace{9pt}Wind                        & \%           & 0              &           & 0             &           & 21.3       &           & 33.7     \\
 \hspace{9pt}Battery charge              & \%           & 0              &           & -0.23         &           & -1.3        &           & -8.7      \\
 \hspace{9pt}Battery discharge           & \%           & 0              &           & 0.19         &           & 1.0        &           & 7.0       \\
 \hspace{9pt}Load curtailment            & \%           & 0.00002              &           & 0.0014            &           & 0.0014            &           & 0.0009  \\
 \hspace{9pt}Grid                        & \%           & 2.4         &           & 3.9       &           & 4.5       &           & 9.5       \\
 \textit{Curtailment}\vspace{3pt}       &               &                &           &               &           &               &           &               \\
\hspace{9pt}PV                          & \%          & 0          &      & 0.9       &          & 1.1        &      & 2.0      \\
\hspace{9pt}Wind                        & \%           & --         &            & --             &           & 7.7       &           & 21.4        \\

\bottomrule
\end{tabular}}
\end{center}\vspace{12pt}
Note: The peak electricity load in 2020 is 573 kW. The total electricity load energy in 2020 is 3,252 MWh. The grid capacity is constrained to 100 kW. 
\end{table}

\begin{figure}[!t]
\centering
\includegraphics[width=.835\linewidth]{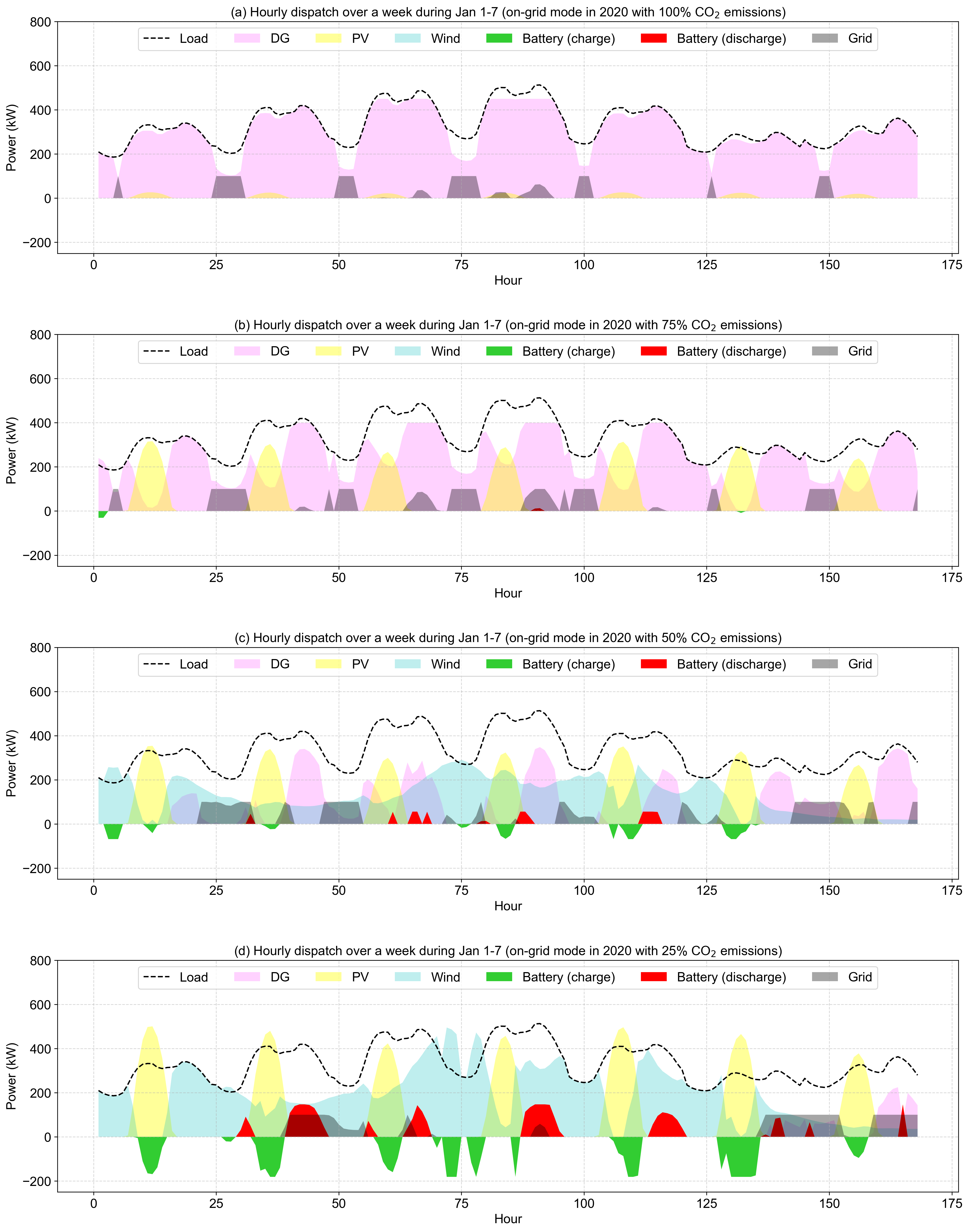}
\caption{Hourly electricity dispatch during January 1-7 for on-grid case in 2020 at different decarbonization levels. The grid capacity is constrained to 100 kW.\\}
\end{figure}

Next, the grid connection is taken into consideration to see how the optimal investment and operation of the generation portfolio change for an on-grid case in 2020.  As shown in \textbf{Table 7} and \textbf{Figure 7}, the changes in total system cost with respect to carbon constraints for the on-grid cases are similar to the results from the off-grid cases. If the unconstrained scenario is considered as the baseline, stricter carbon constraints will increase the total cost by 5.3\%, 17.8\%, and 44.9\% for 75\%, 50\%, and 25\% carbon limits, respectively. Note that the zero-emission case is not added to these results, since it gives the same result as the off-grid counterpart. This is due to the assumption that the grid electricity has non-zero carbon intensity. However, there are noticeable differences between on-grid and off-grid cases in terms of the installed capacities, energy share, and costs. For instance, on average 15\% of the thermal plant capacity is replaced by electricity purchases from the grid. There are slightly less investments in other technologies when grid electricity is available. The total costs of on-grid cases are generally slightly less than those of off-grid cases (see \textbf{Tables 6 and 7}). The average decrease of the total costs from off-grid to on-grid mode is around 2\% in this study. The impact of on-grid mode on the RES penetration rate is not significant, as the grid purchases primarily replace the generation from the thermal plant. Without any carbon constraint, the RES penetration is 2\% and 5\% in on-grid and off-grid modes, going up to 74-79\% under the 75\% CO$_2$ emission constraint. The carbon intensity for the unconstrained on-grid case is 502 g/kWh (slightly higher than for the off-grid counterpart).  

Another observation is that, as the CO$_2$ constraints become more stringent, the thermal plant generation decreases while the purchase from the grid increases. This is because the CO$_2$ emission rate of the thermal plant is assumed larger than that of the grid (see \textbf{Table 2}). In our assumptions, CO$_2$ emissions from the grid are considered to be a constant average value than that of the thermal power plant due to the mix of renewables. However, if an hourly resolution for grid emissions is assumed, there may be hours with grid generation from more polluting plants such as coal and the optimal solution will most likely select dispatching the thermal power plant instead of grid purchase during such periods. Since the thermal plant capacity is lower, the renewable energy and battery storage will become more attractive options to meet the demand during the time when the electricity market price is high. These results illustrate how complex and sensitive the interactions between DER technologies could be, justifying a comprehensive optimization approach for planning these systems embedded in our formulation.

A sample optimal hourly dispatch from each case over January 1-7 (typical winter week) is depicted by \textbf{Figure 8}. Again, similar patterns can be observed that the carbon constraint will promote the use of PV, wind, and ESS and that batteries are often used for load shifting. Another interesting observation is that the grid would usually supply the energy in off-peak hours, when the market price is relatively low. However, since the thermal plant capacity is reduced, the grid supplies will also be adopted to meet the peak demand when the renewable energy and battery storage are not sufficiently available. In general, within the current cost and price assumptions in 2020, the battery ESS is not economically viable at scale in the generation portfolios in both on-grid and off-grid cases, unless strict CO$_2$ emission policies are applied.

\subsection{Battery cost variations in 2050}

A set of scenarios in 2050 is explored to look into the impact of uncertainty in the battery cost projections on the optimal generation portfolio. It is assumed that the availability factors of solar PV and wind energy in 2050 improve due to technological advancements, while the electric load profiles are assumed the same as in 2020 for the sake of comparison. Given the high uncertainty of predicting the electricity market prices in 2050, the on-grid mode is not considered here. The experiments run the model with different battery costs and a no-battery option to analyze the role of battery energy storage in the future distributed power systems, under unconstrained and zero CO$_2$ emission assumptions in 2050. 

\begin{figure}[!b]
\makebox[\textwidth][c]{
\centering
\includegraphics[width=1.0\linewidth]{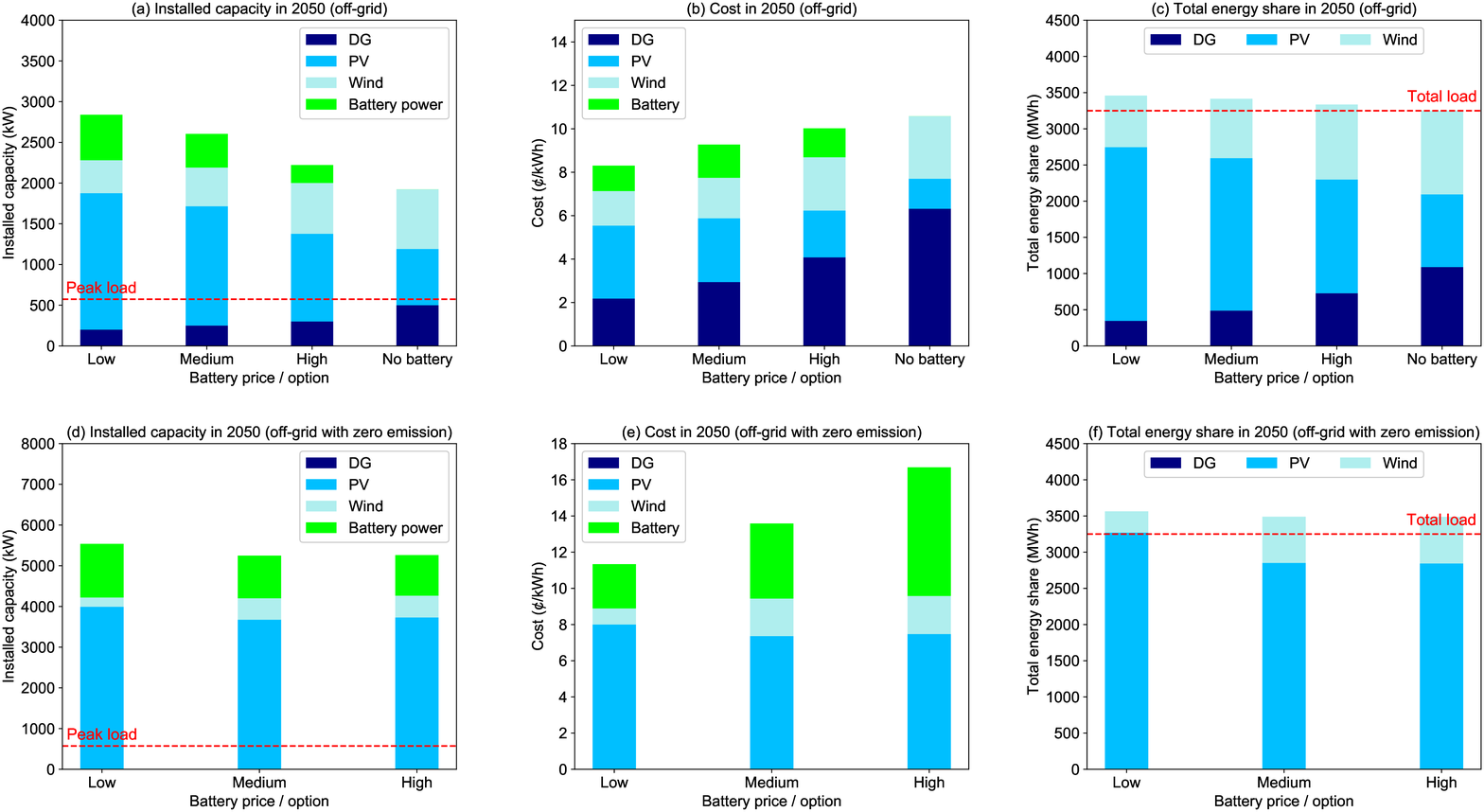}}
\caption{Optimal system configurations and costs for off-grid case in 2050 at different levels of battery option and CO$_2$ emission constraint.}
\end{figure}

As shown in \textbf{Table 8} and \textbf{Figure 9(a)-(c)}, ESS cost reductions help promote the renewable energy and battery storage adoption significantly in 2050 under unconstrained emissions. The no-battery scenario ends up with the highest costs and highest CO$_2$ emissions (10.6 \textcent/kWh and 164 g/kWh), while high levels of renewable energy and batteries are deployed with lower battery costs, leading to decrease in both the energy costs and CO$_2$ emissions. The decrease in the energy costs amounts to 22\% under the lowest battery price case, in which the CO$_2$ emissions decrease 68\%. These results demonstrate that the future costs of ESS may have a large impact on decarbonization, while the impact on the final electricity price is less prominent. Another observation is that, even under the high battery cost in 2050, the resulting generation portfolio yields less CO$_2$ emissions than a 25\%-carbon-constraint scenario in 2020 (see \textbf{Tables 6-8}). The RES share is 66\% without battery and goes up to 89\% under the lowest battery cost. Even with conservatively high battery cost projections for 2050, the RES share goes up to 78\%. In general, the increased battery capacity in combination with PV and wind is an effective carbon reduction measure for the local system, as these technologies allow for increased local renewable generations and decreased thermal plant outputs. These results indicate that ESS will be an increasingly important asset in the future decarbonized power generation portfolio.

\begin{table}[]
\small
\begin{center}
\caption{Detailed investment portfolios for off-grid case in 2050 at different levels of battery option and CO$_2$ emission constraint.}
\makebox[\textwidth][c]{
\begin{tabular}{llllllllll}
\toprule
\textbf{100\% CO$_2$ emissions} & \textbf{Variable}          & \textbf{Unit} & \textbf{Low} & \textbf{} & \textbf{Medium} & \textbf{} & \textbf{High} & \textbf{} & \textbf{No battery} \\ \hline
& \rule{0pt}{3ex}\textit{Installed capacity}\vspace{2pt} &               &              &           &                 &           &               &           &                     \\
& \hspace{9pt}Thermal plant                          & kW            & 200          &           & 250             &           & 300           &           & 500                 \\
& \hspace{9pt}PV                          & kW            & 1,676         &           & 1,465            &           & 1,077          &           & 691                 \\
& \hspace{9pt}Wind                        & kW            & 405          &           & 475             &           & 624           &           & 731                 \\
& \hspace{9pt}Battery power               & kW            & 559          &           & 414             &           & 220           &           & --                   \\
& \hspace{9pt}Battery energy              & kWh           & 8,787         &           & 4,558            &           & 2,114          &           & --                  \vspace{3pt} \\
& \textit{Cost \& emission}\vspace{3pt}   &               &              &           &                 &           &               &           &                     \\
& \hspace{9pt}Total                       & \$            & 269,862      &           & 301,329         &           & 326,020       &           & 344,006             \\
& \hspace{9pt}Thermal plant                          & \$            & 70,889      &           & 95,520          &           & 132,505       &           & 205,290             \\
& \hspace{9pt}PV                          & \$            & 109,176      &           & 95,463          &           & 70,165        &           & 45,042              \\
& \hspace{9pt}Wind                        & \$            & 51,743       &           & 60,700          &           & 79,791        &           & 93,429              \\
& \hspace{9pt}Battery                     & \$            & 38,053       &           & 49,644          &           & 43,208        &           & --                   \\
& \hspace{9pt}Load curtailment            & \$            & 0            &           & 0               &           & 351           &           & 246                 \\
& \hspace{9pt}CO$_2$ emissions            & kg            & 169,485      &            & 298,824         &           & 356,372       &           & 533,257           \vspace{3pt}  \\
& \textit{Energy share}\vspace{3pt}       &               &              &           &                 &           &               &           &                     \\
& \hspace{9pt}Thermal plant               & \%           & 10.7     &           & 15.0         &           & 22.4       &           & 33.5           \\
& \hspace{9pt}PV                          & \%           & 73.9    &           & 64.8       &           & 48.3     &           & 30.9           \\
& \hspace{9pt}Wind                        & \%           & 21.9      &           & 25.2         &           & 31.9     &           & 35.6           \\
& \hspace{9pt}Battery charge              & \%           & -34.4    &           & -26.5         &           & -13.5       &           & --                   \\
& \hspace{9pt}Battery discharge           & \%           & 27.8      &           & 21.4         &           & 11.0       &           & --                   \\
& \hspace{9pt}Load curtailment            & \%           & 0            &           & 0               &           & 0.0008            &           & 0.0006                  \\
& \textit{Curtailment}\vspace{3pt}       &               &                &           &               &           &               &           &               \\
& \hspace{9pt}PV                          & \%          & 5.4          &      & 5.2       &          & 3.8        &      & 4.1      \\
& \hspace{9pt}Wind                        & \%          & 19.3         &     & 20.8        &           & 23.9      &      & 27.5   \\

\specialrule{.1em}{.05em}{.05em} \rule{0pt}{2.5ex}
\textbf{0\% CO$_2$ emissions} & \textbf{Variable}          & \textbf{Unit} & \textbf{Low} & \textbf{} & \textbf{Medium} & \textbf{} & \textbf{High} & \textbf{} & \textbf{No battery} \\ \hline
& \rule{0pt}{3ex}\textit{Installed capacity}\vspace{2pt} &               &              &           &                 &           &               &           &                     \\
& \hspace{9pt}Thermal plant               & kW            & 0             &           & 0             &             & 0           &           & 0                  \\
& \hspace{9pt}PV                          & kW            & 3,992         &           & 3,673            &           & 3,730          &           & 1,155                  \\
& \hspace{9pt}Wind                        & kW            & 227          &           & 527             &           & 535           &           & 7,445                  \\
& \hspace{9pt}Battery power               & kW            & 1,318          &           & 1,048             &           & 997           &           &   --                 \\
& \hspace{9pt}Battery energy              & kWh           & 19,709         &           & 17,425            &           & 17,174          &           &   --                \vspace{3pt} \\
& \textit{Cost \& emission}\vspace{3pt}   &               &              &           &                 &           &               &           &                     \\
& \hspace{9pt}Total                       & \$            & 368,407      &           & 441,541         &           & 542,393       &           & 1,400,625             \\
& \hspace{9pt}Thermal plant                          & \$            & 0      &           & 0          &           & 0       &           &  0            \\
& \hspace{9pt}PV                          & \$            & 260,023      &           & 239,283          &           & 242,938        &           & 75,199               \\
& \hspace{9pt}Wind                        & \$            & 28,956       &           & 67,346          &           & 68,286        &           &949,420               \\
& \hspace{9pt}Battery                     & \$            & 79,428       &           & 134,912          &           & 231,169        &           & --                   \\
& \hspace{9pt}Load curtailment            & \$            & 0            &           & 0               &           & 0           &           & 376,006                 \\
& \hspace{9pt}CO$_2$ emissions            & kg            & 0      &            & 0         &           & 0       &           &  0          \vspace{3pt}  \\
& \textit{Energy share}\vspace{3pt}       &               &              &           &                 &           &               &           &                     \\
& \hspace{9pt}Thermal plant                          & \%           & 0      &           & 0         &           & 0       &           &   0         \\
& \hspace{9pt}PV                          & \%           & 100.7    &           & 87.7       &           & 87.4     &           &  40.6          \\
& \hspace{9pt}Wind                        & \%           & 9.0      &           & 19.7         &           & 19.8     &           & 58.5            \\
& \hspace{9pt}Battery charge              & \%           & -51.3    &           & -38.6         &           & -38.3       &           & --                   \\
& \hspace{9pt}Battery discharge           & \%           & 41.5      &           & 31.3         &           & 31       &           &   --                 \\
& \hspace{9pt}Load curtailment            & \%           & 0            &           & 0               &           & 0            &           & 0.88                  \\
& \textit{Curtailment}\vspace{3pt}       &               &                &           &               &           &               &           &               \\
& \hspace{9pt}PV                          & \%          & 45.9          &      & 48.8       &          & 49.7        &      & 24.6      \\
& \hspace{9pt}Wind                        & \%          & 40.6          &     & 44.4        &           & 44.7      &      & 88.3   \\
\bottomrule
\end{tabular}}
\end{center}\vspace{12pt}
Note: The peak electricity load in 2050 is 573 kW. The total electricity load energy in 2050 is 3,252 MWh.
\end{table}

The study further investigates a zero-emission scenario in 2050 with PV, wind, and different cost assumptions for ESS, as shown in \textbf{Table 8} and \textbf{Figure 9(d)-(f)}. In terms of the installed capacities, eliminating the thermal plant due to the zero-emission restriction leads to substantial investment in ESS. The installed capacity of PV also increases significantly in the zero-emission scenario, while wind capacity does not increase substantially and even decreases in some cases. These changes in the generation portfolio lead to a 37-66\% increase in the electricity costs for different battery cost projections compared to the case with no restriction on CO$_2$ emissions. Another important impact of obtaining zero-emission system is the significantly higher curtailment rate for variable renewable energy, with PV and wind curtailments reaching almost 50\% in the high battery price case. A zero-emission case without battery is also examined to explore the feasibility of meeting demand with PV and wind only. In this case, investment in wind is preferred over PV. The total system cost and load curtailment increase substantially, indicating that a zero-emission system without ESS seems not feasible from both economic and reliability perspectives under the assumptions used in this study.

A sample optimal hourly generation dispatch from the cases without CO$_2$ emission constraints over January 1-7 (typical winter week) is depicted in \textbf{Figure 10}. Owing to the high penetration of renewable energy and ESS, the dispatch of the thermal plant in 2050 is significantly reduced compared to 2020. Besides, it can be seen that the battery is fully cycled on most days (i.e., ESS is largely replacing the dispatchability provided to the system from the thermal plant in 2020). In predicting the optimal portfolio of resources for low-carbon distributed power systems, it would be remiss not to consider ESS given the substantial techno-economic benefits in the future. Therefore, even with the degradation and shortened life concerns, given the projected technology advancements and falling prices, battery technologies exhibit an attractive option and should be actively considered in the generation portfolio for distributed and decarbonized power systems. However, to obtain more accurate economic estimations of the battery value in power system planning and operation, a more detailed battery representations including the degradation impacts should be incorporated into the expansion models.\\

\begin{figure}[!t]
\centering
\includegraphics[width=.835\linewidth]{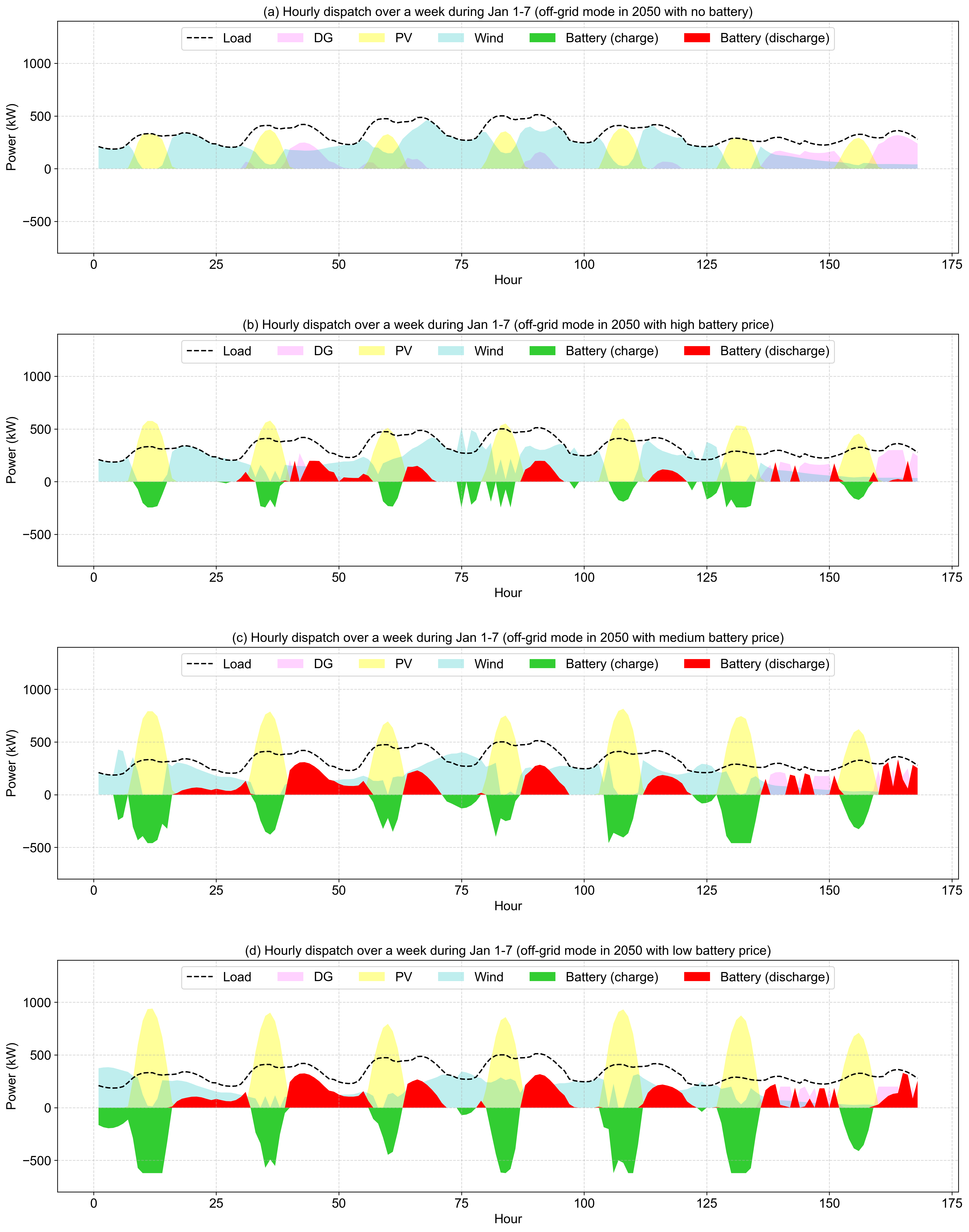}
\caption{Hourly electricity dispatch during January 1-7 for off-grid case in 2050 at different levels of battery option without CO$_2$ emission constraint.}
\end{figure}

\subsection{Transition from 2020 to 2050}

\begin{table}[!t]
\small
\begin{center}
\caption{Detailed investment portfolios for off-grid case without CO$_2$ emission constraint during transition from 2020 to 2050.}
\begin{tabular}{lllllllll}
\toprule
\textbf{Variable}          & \textbf{Unit} & \textbf{2020} & \textbf{} & \textbf{2030} & \textbf{} & \textbf{2040} & \textbf{} & \textbf{2050} \\ \hline
\rule{0pt}{3ex}\textit{Installed capacity}\vspace{2pt} &               &                &           &               &           &                 \\
\hspace{9pt}Thermal plant                          & kW            & 500            &           & 400           &           & 300          &           & 250         \\
\hspace{9pt}PV                          & kW            & 143            &           & 650        &              & 1,127          &            & 1,466       \\
\hspace{9pt}Wind                        & kW            & 0              &           & 303             &           & 490           &           & 475     \\
\hspace{9pt}Battery power               & kW            & 31             &           & 126            &           & 243             &      & 414      \\
\hspace{9pt}Battery energy              & kWh           & 157            &           & 799           &       & 2,486        &           & 4,558         \vspace{3pt}  \\
\textit{Cost \& emission}\vspace{3pt}   &               &                &           &               &           &                            \\
\hspace{9pt}Total                       & \$           & 277,782         &       & 325,247        &       & 318,875        &           & 301,330        \\
\hspace{9pt}Thermal plant                          & \$           & 258,945         &           & 203,667        &      & 133,705        &           & 95,521        \\
\hspace{9pt}PV                          & \$           & 14,308          &          & 59,947         &         & 83,284       &          & 95,464        \\
\hspace{9pt}Wind                        & \$           & 0              &           & 47,396             &           & 67,108         &      & 60,701        \\
\hspace{9pt}Battery                     & \$           & 4,131       &           & 14,049          &           & 34,778        &           & 49,644         \\
\hspace{9pt}Load curtailment            & \$           & 398            &           & 189          &           & 0           &           & 0         \\
\hspace{9pt}CO$_2$ emissions            & kg            & 1,589,352        &           & 911,440       &           & 441,880        &           & 238,824        \vspace{3pt}  \\
\textit{Energy share}\vspace{3pt}       &               &                &           &               &           &                    \\
\hspace{9pt}Thermal plant               & \%           & 94.5    &           & 55.2      &       & 27.3       &           & 15.0      \\
\hspace{9pt}PV                          & \%          & 5.5          &      & 26.8        &          & 48.4        &      & 64.8  \\
\hspace{9pt}Wind                        & \%           & 0         &            & 18.4             &           & 27.1       &           & 25.2      \\
\hspace{9pt}Battery charge              & \%           & -0.06      &           & -1.79         &           & -14.37             &     & -26.47     \\
\hspace{9pt}Battery discharge           & \%           & 0.05    &           & 1.45         &           & 11.64             &       & 21.44     \\
\hspace{9pt}Load curtailment            & \%           & 0.0009             &           & 0.0004           &           & 0            &           & 0             \\
\textit{Curtailment}\vspace{3pt}       &               &                &           &               &           &                          \\
\hspace{9pt}PV                          & \%          & 0          &      & 0.06        &          & 2.27        &      & 5.16      \\
\hspace{9pt}Wind                        & \%           & --         &            & 1.86             &           & 14.13       &           & 20.81      \\

\bottomrule
\end{tabular}
\end{center}\vspace{12pt}
Note: The peak electricity load is 573 kW. The total electricity load energy is 3,252 MWh.\\
\end{table}

\begin{figure}[!t]
\makebox[\textwidth][c]{
\centering
\includegraphics[width=1.0\linewidth]{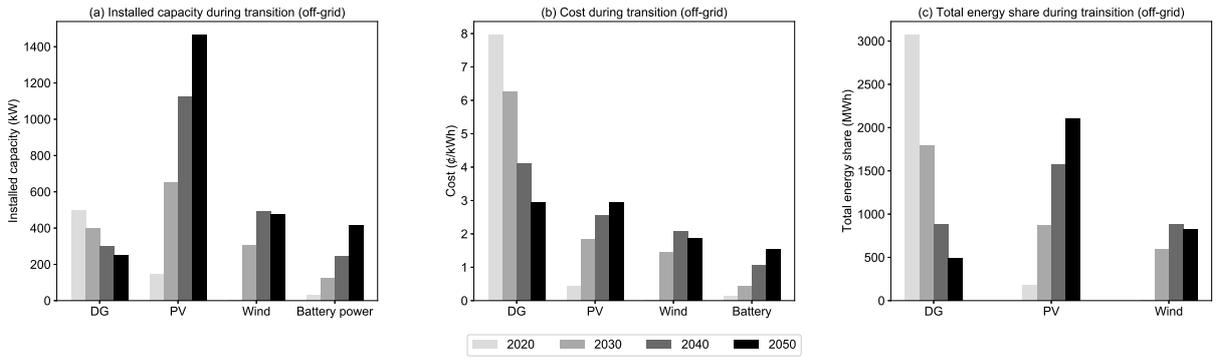}}
\caption{Optimal system configurations and costs for off-grid case without CO$_2$ emission constraint during transition from 2020 to 2050.}
\end{figure}

Finally, to show an optimal transition from now to mid-century, the off-grid scenarios without CO$_2$ emission constraint are performed for 2020, 2030, 2040, and 2050. The results are shown in \textbf{Table 9} and \textbf{Figure 11}. These results indicate that technology cost reductions toward 2050 will motivate higher renewables and storage investments, reducing the share of the thermal power plant and the CO$_2$ emissions (from 489 g/kWh in 2020 to 280, 136, and 73 g/kWh in 2030, 2040, and 2050, respectively). Even without any CO$_2$ emission constraint, the share of renewables in supplying the load reaches 85\% in 2050. This transition requires ESS with longer durations, such that it increases from 5 hours in 2020 to 6.5, 10, and 11 hours in 2030, 2040, and 2050, respectively. Another observation is that the optimal energy cost per MWh increases in 2030 (to \$100 from \$85 in 2020) and then decreases in 2040 (\$98) and 2050 (\$93), mostly due to lower investment and operation costs of the thermal power plant.

\section{Conclusion}

Energy storage for distributed power systems is currently a vibrant research area. As battery technologies mature with falling prices, their attractiveness to an ever-growing spectrum of decarbonized applications is set to flourish. In this expansive phase, research evolves at a rapid pace producing a plethora of models and analyses, with substantial efforts on systematization of optimally sizing storage resources in conjunction with other DERs. For the field to move further, certain aspects should be detailed and pursued with higher priority.

This paper proposes an advanced formalization of the battery ESS model and its integration into a distributed system planning problem of optimally sizing generation resources. An MILP formulation of the planning problem that accounts for the dynamic efficiencies, power limits, and degradation effects of a Li-ion battery system is presented under a unified framework to analyze the role of energy storage in future distributed power systems. Moreover, a single-node case study is conducted to empirically explore the established program at different model complexities and parametric scenarios.

The first discovery is that detailed characterizations of ESS dynamic efficiencies and power limits, while requiring substantial computational cost, might not significantly impact the optimal generation portfolio in a planning problem. Given the trade-off between computational cost and model performance, our results indicate that it is not worthwhile to consider charge/discharge efficiency and maximum power dynamic properties in the battery system modeling for the planning problem. On the other hand, the effect of battery degradation is an important aspect in the optimal generation portfolio for power system and will produce substantially different outcomes. A failure to incorporate degradation effects will likely generate suboptimal solutions leading to unrealistic estimations of the ESS's potential benefits. 

A second observation from our case study is that, with the current technology costs (i.e., in 2020), battery storage does not seem very attractive in the resource portfolio, especially when the system is connected to the main grid. Considering the high technology costs, battery degradation impact, and low capacity factor for renewables, the economically optimal solution is to invest in the thermal power plant in 2020. We find that only under strict CO$_2$ emission constraints will the use of renewable energy and battery ESS be promoted to allow for more local clean generation and corresponding reductions in thermal generation and grid purchases of electricity. The final electricity cost increases almost 50\%, if we choose to adopt a large portion of renewable energy and battery energy storage to reduce emissions by 75\% in 2020.

However, our final predictive investigation emphasizes that energy storage will be considered as an increasingly important asset in future distributed power systems on the pathway toward a low (or even zero) carbon future. Rapidly decreasing battery costs  will enable us to benefit from the combined values of the renewable energy and battery technologies. Even with high battery cost projections, the share of renewables and the level of decarbonization increases substantially in the optimal solutions for 2050 (e.g., 11\% higher RES penetration and 33\% lower CO$_2$ emissions for high battery cost projection scenario).  Therefore, increased battery usage in combination with PV and wind presents an important deep decarbonization measure for future distributed power systems. Overall, the results indicate that, with improved technologies and decreased costs of batteries and renewables, the optimal solution will be investing in a battery-renewable combination instead of thermal power plant. Battery ESS will exhibit an attractive option in the generation portfolio to promote high penetration of clean energy in 2050.

It is important to note that the presented results in this study are obtained under the mentioned cost assumptions and should be interpreted accordingly.  Also, several dimensions of this study should be further explored in the future. One direction is to integrate more flexible DER options and system components into the case study and see how the battery storage interacts with other flexible assets. Another suggestion would be to extend the current formulation into a multi-node system and consider interaction between ESS and network design. Nevertheless, the formalized planning model developed in this study contributes to a more appropriate assessment of energy storage in meeting local electricity demand and can be smoothly scaled up to more complex systems.

\section*{Acknowledgment}

This work was supported by ExxonMobil through its membership in the MIT Energy Initiative. Executions of the optimization program have benefited from the MIT Lincoln Laboratory Supercomputing Center. We thank Rupamathi Jaddivada (MIT), Yibao Jiang (Zhejiang University), and Lauren Milechin (MIT) for their generous help during the development of this work.

\printbibliography

\end{document}